\documentclass[fleqn,usenatbib]{mnras}

\usepackage{newtxtext,newtxmath}

\usepackage[T1]{fontenc}

\DeclareRobustCommand{\VAN}[3]{#2}
\let\VANthebibliography\thebibliography
\def\thebibliography{\DeclareRobustCommand{\VAN}[3]{##3}\VANthebibliography}

\usepackage{graphicx}	
\usepackage{amsmath}	

\usepackage{subcaption}
\usepackage{caption}
\usepackage{multirow} 
\usepackage{float} 
\usepackage{natbib}
\usepackage{hyperref}
\defcitealias{chandro-gomez_inprep}{Paper~I}

\title[MQGs at $z\ge2$ in COLIBRE -- II]{Unveiling the population of massive quenched galaxies at $z\ge2$ in the COLIBRE simulations -- II. The role of AGN feedback and environment on their emergence}

\author[\'Angel Chandro-G\'omez et al.]{
\parbox[t]{\textwidth}{
\vspace{-0.5cm}
\'Angel Chandro-G\'omez$^{1,2}$\thanks{E-mail: angel.chandrogomez@research.uwa.edu.au}, 
Claudia del P. Lagos$^{1,2}$,
Chris Power$^{1,2}$,
\textcolor{black}{William} M. Baker$^{3}$,
Alejandro Benítez-Llambay$^{4}$,
Evgenii Chaikin$^{5,6}$,
Harry G. Chittenden$^{7,8}$,
Camila Correa$^{6}$,
Carlos S. Frenk$^{5}$,
Filip Huško$^{6}$,
\textcolor{black}{Kei Ito}$^{9,10}$,
Robert J. McGibbon$^{6}$,
Themiya Nanayakkara$^{7,8}$,
Sylvia Ploeckinger$^{11}$,
Alexander J. Richings$^{12,13}$,
Matthieu Schaller$^{6,14}$,
Joop Schaye$^{6}$,
James W. Trayford$^{15}$
and \textcolor{black}{Francesco Valentino}$^{9,10}$}
\vspace*{8pt} \\
$^{1}$ International Centre for Radio Astronomy Research (ICRAR), The University of Western Australia, 35 Stirling Highway, Crawley, WA 6009, Australia\\
$^{2}$ ARC Centre for All-Sky Astrophysics in 3 Dimensions (ASTRO 3D), Australia\\
$^{3}$ DARK, Niels Bohr Institute, University of Copenhagen, Jagtvej 155A, DK-2200 Copenhagen, Denmark\\
$^{4}$ Dipartimento di Fisica G. Occhialini, Universit\`a degli Studi di Milano Bicocca, Piazza della Scienza, 3 I-20126 Milano MI, Italy\\
$^{5}$ Institute for Computational Cosmology, Department of Physics, University of Durham, South Road, Durham, DH1 3LE, UK\\
$^{6}$ Leiden Observatory, Leiden University, PO Box 9513, NL-2300 RA Leiden, The Netherlands\\
$^{7}$ Centre for Astrophysics and Supercomputing, Swinburne University of Technology, P.O. Box 218, Hawthorn VIC 3122, Melbourne, Australia\\
$^{8}$ \textit{JWST} Australia Data Centre, Swinburne Advanced Manufacturing and Design Centre, John Street, Hawthorn VIC 3122, Melbourne, Australia\\
$^{9}$ \textcolor{black}{Cosmic Dawn Center (DAWN), Copenhagen, Denmark}\\
$^{10}$ \textcolor{black}{DTU Space, Technical University of Denmark, Elektrovej 327, DK2800 Kgs. Lyngby, Denmark}\\
$^{11}$ Department of Astrophysics, University of Vienna, Türkenschanzstrasse 17, A-1180 Vienna, Austria\\
$^{12}$ Centre for Data Science, Artificial Intelligence and Modelling, University of Hull, Cottingham Road, Hull, HU6 7RX, UK\\
$^{13}$ E. A. Milne Centre for Astrophysics, University of Hull, Cottingham Road, Hull, HU6 7RX, UK\\
$^{14}$ Lorentz Institute for Theoretical Physics, Leiden University, PO Box 9506, NL-2300 RA Leiden, The Netherlands\\
$^{15}$ Institute of Cosmology and Gravitation, University of Portsmouth, Dennis Sciama Building, Burnaby Road, Portsmouth PO1 3FX, UK\\
\vspace*{-0.5cm}}

\date{Accepted XXX. Received YYY; in original form ZZZ}

\pubyear{2025}

\begin{document}
\label{firstpage}
\pagerange{\pageref{firstpage}--\pageref{lastpage}}
\maketitle

\begin{abstract}
\textcolor{black}{Early ($z \geq 2$) Massive ($M_{\star} > 10^{10}\,\mathrm{M_{\odot}}$), Quenched Galaxies (MQGs) challenge current galaxy formation models}. In this \textcolor{black}{series}, we \textcolor{black}{study these systems using} the new {\sc COLIBRE} cosmological hydrodynamical simulations. \textcolor{black}{Following the broad agreement between their predictions and observations found in the first paper, this second paper explores} the processes driving galaxies to become massive and quenched in {\sc COLIBRE}, identifying Active Galactic Nucleus (AGN) feedback as the primary quenching mechanism \textcolor{black}{in both the thermal (L200m6 simulation) and hybrid (thermal+jet, L200m7h simulation) AGN feedback models implemented. However, the two models behave differently: while the thermal model efficiently quenches massive galaxies at $z>3$, the hybrid model is less effective because black holes (BHs) grow more slowly in the early Universe, and the jet component, which dominates the feedback energy, acts on longer timescales to impact galaxies. Both models predict quasar-like MQGs (AGN with $L_{\rm bol}\gtrsim10^{45}\,\mathrm{erg\,s^{-1}}$), with the most luminous systems associated with more recently quenched galaxies.} Compared to star-forming galaxies of similar mass, MQGs host more massive BHs and exhibit higher star formation efficiencies. These differences arise \textcolor{black}{primarily} from their environments \textcolor{black}{before quenching}, particularly at local ($\rm 0.3\,cMpc$) to intermediate scales ($\rm 1.0\,cMpc$), where overdense regions are associated with enhanced gas inflows, higher BH accretion and, hence, feedback power. We find that about \textcolor{black}{$54\%$ ($20\%$)} of the $z=3$ MQGs survive as the main progenitors of $z=0$ galaxies, although up to \textcolor{black}{$56\%$ ($60\%$)} experience rejuvenation episodes \textcolor{black}{at a given redshift $z<3$ in L200m6 (L200m7h)}. \textcolor{black}{Our results highlight the central role of BH growth, AGN feedback and environment in driving rapid quenching in the early Universe.}
\end{abstract}




\section{Introduction}
\label{sec:intro} 

\textcolor{black}{The population of massive ($M_{\star} > 10^{10}\,\mathrm{M}_{\odot}$), quenched (negligible ongoing star formation) galaxies (hereafter MQGs) at high redshift ($z \gtrsim 2$) uncovered by the \textit{James Webb Space Telescope} (\textit{JWST}) is proving to be far more abundant than previously inferred \citep{glazebrook17, schreiber18, valentino20, weaver22}. A growing number of studies measuring their number densities \citep{nanayakkara22, carnall23, baker25, stevenson25} have revealed a substantial population of MQGs already in place in the early Universe, including some extreme examples \textcolor{black}{such as a galaxy} observed at $z=7.3$ \citep{weibel25} and \textcolor{black}{a system} whose stellar populations may have formed as early as $z\approx11$ \citep{glazebrook24}. These systems challenge $\Lambda$ Cold Dark Matter ($\Lambda$CDM) galaxy formation scenarios, as their rapid stellar mass assembly sometimes requires extremely efficient early star formation \citep{carnall24, glazebrook24, degraaff25}.}

\textcolor{black}{Their rapid quenching presents an additional challenge, poorly understood yet}. Given the short quenching timescales of MQGs, the most plausible mechanism is seen as active galactic nuclei (AGN) feedback, the injection of energy and momentum by central supermassive black holes (BHs) into their surroundings. AGN feedback is understood to play a crucial role in shaping the properties of massive galaxies at low redshifts, being necessary to explain the high-mass end of the stellar mass function (SMF) \citep[e.g.][]{benson03}. It suppresses star formation by expelling cold gas through powerful outflows (ejective feedback; \citealt{dimatteo05, booth&schaye09}) and by heating the surrounding medium and preventing further accretion (preventive feedback; \citealt{bower06, croton06}). Observational evidence indicates that AGNs are active in early MQGs, through detections of X-rays \citep{ito22, stevenson25, baker25b}, broad emission lines \citep{martinez-marin24} and AGN-driven outflows seen in absorption lines \citep{davies24, d'eugenio24, park24}.

\textcolor{black}{Modern galaxy formation and evolution models \citep[see e.g.][for reviews]{somerville15, crain23} provide an ideal framework to investigate the role of AGN feedback in quenching massive galaxies at such high redshift. These models generally calibrate their sub-grid physics prescriptions to reproduce galaxy populations in the local Universe \citep[e.g.][]{crain15,chaikin25}, making their predictions at high redshift particularly valuable tests of the underlying physics. In essentially all models, AGN feedback is the primary mechanism invoked to halt star formation in massive systems \citep{benson03, bower06, croton06}. However, the implementation of AGN feedback differs substantially between models. Some of them inject the feedback energy thermally, others kinetically, while some combine multiple feedback modes that operate under different conditions. Important differences also arise from BH growth modelling such as the treatment of gas accretion onto the BH \citep{bondi52}, including whether super-Eddington accretion is allowed, which may be particularly relevant in the early Universe \citep{husko25b}.} 

\textcolor{black}{For example, among cosmological hydrodynamical simulations, {\sc IllustrisTNG} \citep{pilepich18} implements thermal and kinetic AGN feedback modes, \textcolor{black}{with the transition between them determined by whether the BH accretion rate (BHAR) is above or below a BH-mass-dependent threshold. BH growth is limited by the Eddington accretion and is fuelled only by hot-gas accretion \citep{bondi52}.} {\sc Magneticum} \citep{steinborn15} includes both hot- and cold-gas accretion channels, together with two thermal feedback modes (quasar and radio) operating with different efficiencies, but otherwise modelled in the same way. {\sc Simba} \citep{dave19} incorporates torque-limited cold-gas accretion in addition to hot-gas accretion, allows for super-Eddington accretion \textcolor{black}{up to three times the Eddington accretion rate}, and combines kinetic and thermal feedback whose relative importance depends on both the accretion rate and BH mass. {\sc EAGLE} \citep{schaye15}, by contrast, uses a thermal-only AGN feedback prescription with only hot-gas accretion and without super-Eddington accretion. Implementation differences in the BH growth and AGN feedback prescriptions can strongly affect the predicted abundance and properties of MQGs \citep{lagos25}.}

\textcolor{black}{Furthermore, galaxy formation models also make it possible to investigate the environments in which MQGs reside and how these environments may influence their formation and quenching. Observationally, the first indications are emerging that MQGs at high-$z$ tend to inhabit dense environments \citep[e.g.][]{jin24, ito25b, baker25c, mcconachie25, kakimoto26}. Environment is closely linked to AGN activity at low redshift, where MQGs are preferentially found in dense regions \citep[e.g.][]{kauffmann04}. A similar connection may therefore exist at high redshift, where dense environments could promote the rapid growth of BHs required to quench MQGs efficiently.} 

\textcolor{black}{In this paper, we present the second part of a series focused on understanding the properties and origins of MQGs in the new {\sc COLIBRE} cosmological hydrodynamical simulations \citep{schaye25, chaikin25}, with both papers intended to be read together. The first paper in the series, \citet{chandro-gomez_inprep} (hereafter \citetalias{chandro-gomez_inprep}), investigated the population demographics of MQGs at $z\ge 2$. That work showed that {\sc COLIBRE} reproduces the observed number densities \citep[see also][]{chaikin25b} and SMFs of MQGs in broad agreement with current observational estimates, unlike many previous simulations evolved down to $z=0$ \citep{lagos25}. Accounting for observational and theoretical systematic uncertainties significantly reduces the tension between simulations and observations. \citetalias{chandro-gomez_inprep} also explored other MQG properties, including features in their star formation histories (SFHs), molecular gas and dust fractions, sizes and kinematics, finding results broadly consistent with \textit{JWST} observations and supporting a scenario in which quenching occurs before significant morphological transformation takes place.}

\textcolor{black}{Building on the results of \citetalias{chandro-gomez_inprep}, this paper focuses on developing a deeper understanding of the physical drivers that give rise to MQGs in the {\sc COLIBRE} model. Our main goal is to identify the mechanisms responsible for their rapid formation and quenching. 
Beyond reproducing observed galaxy populations, cosmological simulations, \textcolor{black}{despite limitations in snapshot cadence}, provide temporal information for galaxies that cannot be obtained observationally. 
By tracing the evolutionary histories of individual MQGs backwards in time, we investigate the role of BH growth, AGN feedback, and environment in driving quenching. We also follow these galaxies down to $z=0$ to study their long-term evolution and the possibility of rejuvenation episodes that restart star formation after quenching.}


\textcolor{black}{The structure of this second paper is as follows. In \S~\ref{sec:colibre} we describe the {\sc COLIBRE} simulations, focusing on the BH growth and AGN feedback modelling. \S~\ref{sec:prop} outlines our sample selection and the properties analysed throughout this paper. In \S~\ref{sec:res2}, we follow the evolutionary journey of MQGs predicted by the simulations, examining the mechanisms that quench them, tracing their past evolution to understand the quenching origin and exploring the potential for renewed star formation (rejuvenation). Finally, \S~\ref{sec:conclusions} summarises our main conclusions.}

\section{The COLIBRE simulation suite}
\label{sec:colibre}

\begin{table*}
	\centering
	\caption{{\sc COLIBRE} hydrodynamical simulations used in this work. $L$: the periodic box size in comoving Mpc (cMpc); $N_{\rm b}$: the initial number of baryonic particles in the volume; $N_{\rm DM}$: the number of DM particles in the volume; $m_{\rm g}$: the mean initial gas particle mass in M$_{\odot}$; $m_{\rm DM}$: the mean DM particle mass in M$_{\odot}$; $\epsilon_{\rm com}$: the gravitational softening length in ckpc; $\epsilon_{\rm prop}$: the maximum gravitational softening length in proper kpc (pkpc); and the type of AGN feedback implemented (thermal or hybrid including thermal+jet); $N_{\mathrm{MQ}}(z=3)$: the number of MQGs selected at $z=3$ using the criteria defined in \S~\ref{ssec:prop-mq-def}. We highlight in bold the simulation adopted as fiducial in this work. \textcolor{black}{The L050 boxes are only used in \S~\ref{sssec:res2-agn-feedback-ndens}.}}
	\label{tab:runs}
	\begin{tabular}{cccccccccc}
		\hline
		{\sc COLIBRE} sim. & $L$/cMpc & $N_{\rm b}$ & $N_{\rm DM}$ & $m_{\rm g}$/M$_{\odot}$ & $m_{\rm DM}$/M$_{\odot}$ & $\epsilon_{\rm com}$/ckpc & $\epsilon_{\rm prop}$/pkpc & AGN feedback model & $N_{\mathrm{MQ}}(z=3)$\\
		\hline
        \textbf{L200m6} & $\boldsymbol{200}$ & $\boldsymbol{3008^3}$ & $\boldsymbol{4\times3008^3}$ & $\boldsymbol{1.84\times10^6}$ & $\boldsymbol{2.42\times10^6}$ & $\boldsymbol{1.8}$ & $\boldsymbol{0.7}$ & \textbf{thermal} & $\boldsymbol{470}$\\
        L200m7 & 200 & 1504$^3$ & $4\times1504^3$ & 1.47$\times$10$^7$ & 1.94$\times$10$^7$ & 3.6 & 1.4 & thermal & 1527\\
        L200m7h & 200 & 1504$^3$ & $4\times1504^3$ & 1.47$\times$10$^7$ & 1.94$\times$10$^7$ & 3.6 & 1.4 & hybrid (thermal+jet) & 310\\
        \hline
        \textcolor{black}{L050m6} & 50 & 752$^3$ & $4\times752^3$ & 1.84$\times$10$^6$ & 2.42$\times$10$^6$ & 1.8 & 0.7 & thermal & 5\\
        \textcolor{black}{L050m7} & 50 & 376$^3$ & $4\times376^3$ & 1.47$\times$10$^7$ & 1.94$\times$10$^7$ & 3.6 & 1.4 & thermal & 18\\
        \textcolor{black}{L050m6h} & 50 & 752$^3$ & $4\times752^3$ & 1.84$\times$10$^6$ & 2.42$\times$10$^6$ & 1.8 & 0.7 & hybrid (thermal+jet) & 3\\
        \textcolor{black}{L050m7h} & 50 & 376$^3$ & $4\times376^3$ & 1.47$\times$10$^7$ & 1.94$\times$10$^7$ & 3.6 & 1.4 & hybrid (thermal+jet) & 0\\
        \textcolor{black}{L050m6noagn} & 50 & 752$^3$ & $4\times752^3$ & 1.84$\times$10$^6$ & 2.42$\times$10$^6$ & 1.8 & 0.7 & no AGN  & 0\\
        \textcolor{black}{L050m7noagn} & 50 & 376$^3$ & $4\times376^3$ & 1.47$\times$10$^7$ & 1.94$\times$10$^7$ & 3.6 & 1.4 & no AGN & 0\\
		\hline
	\end{tabular}
\end{table*}

\textcolor{black}{We use the new {\sc COLIBRE}\footnote{\url{https://colibre-simulations.org/}} suite of large-volume cosmological hydrodynamical simulations \citep{schaye25, chaikin25} to carry out this analysis. Relative to other large-volume hydrodynamical simulations, {\sc COLIBRE} incorporates several advances in its sub-grid modelling, which represent physical processes occurring below the simulation’s spatial resolution. A comprehensive description is given in \citet{schaye25} and the individual papers for each sub-grid model. These include: (i) detailed modelling of chemical processes in the cold ISM, with non-equilibrium chemistry for hydrogen and helium enabling gas to cool down to $\approx 10$~K \citep{ploeckinger25}; (ii) self-consistent tracking of dust formation and evolution \citep{trayford25}; (iii) a star formation prescription based on a local gravitational instability criterion \citep{nobels24}; (iv) an updated treatment of chemical enrichment, including turbulent diffusion \citep{correa26}; (v) early stellar feedback \citep{benitez-llambay25} together with revised thermal and kinetic core collapse supernova (CCSN) feedback \citep{chaikin23}; (vi) supermassive BH growth that includes repositioning driven by dynamical friction \citep{bahe22} and allows for super-Eddington accretion; and (vii) AGN feedback implemented either as a purely thermal model (fiducial) or as a hybrid scheme combining thermal energy input (associated with winds and radiation) with kinetic jets regulated by BH spin, which also sets their orientation and efficiency \citep{husko25}. Additional details on (vi) and (vii) are provided in \S~\ref{ssec:colibre-agn}. Model parameters are calibrated \citep{chaikin25, husko25} to reproduce key $z=0$ observables, including the galaxy SMF \citep{driver22}, galaxy sizes \citep{hardwick22}, and BH scaling relations \citep{graham23}. \textcolor{black}{For the hybrid AGN feedback model, the $z=0.2$ bolometric AGN luminosity function \citep{shen20} is additionally included as a calibration constraint.} Each combination of resolution and AGN feedback model is calibrated independently \citep[see table~1 in][]{schaye25}.}

\textcolor{black}{Initial conditions (ICs) for the simulations are generated with {\sc MONOFONIC} \citep{hahn20, michaux21}, and the simulations are evolved using the {\sc SWIFT} code \citep{schaller24}, which employs smoothed particle hydrodynamics (SPH). To minimise spurious energy exchange between dark matter (DM) and stellar particles, the simulations use a DM particle number that is four times larger than that of baryons \citep{ludlow19}. The runs extend from $z=63$ to $z=0$, adopting the DES Y3 ‘3x2pt + All Ext.’ $\Lambda$CDM cosmology \citep{abbott22}, with parameters $h=0.681$, $\Omega_{\Lambda}=0.693922$, $\Omega_{\rm m}=0.306$, $\Omega_{\rm CDM}=0.256011$, and $\Omega_{\rm b}=0.0486$. Snapshots are stored over the redshift range $z=30$ to $z=0$. Haloes are first identified with a friends-of-friends (FoF) algorithm that links DM particles separated by less than 0.2 times the mean inter-particle distance, with baryonic particles attached to the nearest DM neighbour. Subhaloes are then identified using the {\sc HBT-HERONS} code \citep{forouhar25}, which includes an explicit temporal tracking that improves the detection of substructures near halo centres \citep{forouhar25} and prevents unphysical mass fluctuations in merger trees \citep{chandro-gomez25}. 
%
Finally, halo, subhalo and galaxy properties are computed with the {\sc SOAP} code \citep{mcgibbon25}, using a range of 3D and projected apertures. Further details on the simulation methods can be found in \citetalias{chandro-gomez_inprep} and  \citet{schaye25}.}

\textcolor{black}{The {\sc COLIBRE} suite spans a range of simulation volumes (periodic boxes of $L=25-400\rm\,cMpc$ per side), mass resolutions (with mean initial gas and DM particle mass, $m_{\rm g}$ and $m_{\rm DM}$, $\sim10^5-10^7\,\mathrm{M}_{\odot}$), and AGN feedback models (either fiducial thermal or hybrid thermal+jet). In \citetalias{chandro-gomez_inprep}, we adopted the L200m6 run as the fiducial model, as it provides a suitable balance between volume ($L=200\,\rm cMpc$), large enough to sample rare MQGs, and resolution ($m_{\rm g}$ and $m_{\rm DM} \sim 10^6\,\mathrm{M}{\odot}$), sufficient to resolve their internal properties. Its relatively small gravitational softening lengths ($\epsilon_{\rm com}$ and $\epsilon_{\rm prop}$) are particularly advantageous for robust measurements of galaxy sizes and kinematics. That run provided MQG number densities and SMFs in broad agreement with {\sc JWST} observations, as shown in secs.~4.1~and~4.2 of \citetalias{chandro-gomez_inprep}. In this work, we compare this run with L200m7h to assess the impact of the hybrid AGN feedback model. To disentangle resolution effects, we also include L200m7 (the thermal-feedback counterpart at the same resolution) as a reference. Table~\ref{tab:runs} summarises the main characteristics of these simulations. This table also includes several $L=50\,\rm cMpc$ runs with the two AGN feedback models, plus variations without AGN feedback included, which are used only in \S~\ref{sssec:res2-agn-feedback-ndens}. }

\subsection{Supermassive black hole growth \& AGN feedback modelling}
\label{ssec:colibre-agn} 

\textcolor{black}{In {\sc COLIBRE},} BH particles are seeded whenever a non-seeded FoF group exceeds a threshold mass of $M_{\rm FoF,seed}$, at redshifts later than $z=19$. The default $M_{\rm FoF,seed}$ is $\rm 5\times10^{10}\,M_{\odot}$ for the m7 resolution and $\rm 1\times10^{10}M_{\odot}$ for m6 and m5 resolutions. If seeding is needed, the densest gas particle in the halo is converted into a collisionless BH particle with initial mass $M_{\rm BH}=m_{\rm BH,seed}$, typically of order $\rm\sim10^{4}-10^{5}\,M_{\odot}$, following the “heavy seed” scenario in which BHs form via direct collapse of gas clouds or rapid mergers of stars/low-mass BHs \citep[e.g.][]{regan24}. The value depends on the simulation, as listed in table~1 of \citet{schaye25}, since it is a free parameter in the calibration process. BH particles are repositioned according to the prescription in \citet{bahe22} to account for unresolved dynamical friction, ensuring the BH moves towards the galaxy's potential centre.

Each BH particle has two distinct masses associated with it: a sub-grid mass, $M_{\rm BH}$, which governs BH-related physics; and a particle mass for gravity calculations, $m_{\rm BH,part}$ (initially equal to the mass of the gas particle converted into the BH). After seeding, $M_{\rm BH}$ grows through both mergers and gas accretion. \textcolor{black}{While $M_{\rm BH} < m_{\rm BH,part}$, the increase in the sub-grid mass through gas accretion is not accompanied by particle mass growth. Instead, the particle mass is adjusted to account for the fraction of the accreted mass converted into feedback energy. Once $M_{\rm BH} > m_{\rm BH,part}$, the BH particle increases its mass by stochastically "nibbling" mass from neighbouring gas particles, ensuring consistency between the sub-grid and gravitational masses while conserving mass and energy \citep{bahe22}.}

BH mergers occur when two BHs meet the following conditions: their separation is smaller than 3 times the gravitational softening length, the less massive BH lies within the smoothing length of the more massive one, defined with the same kernel used in the SPH; and their relative velocity satisfies $\Delta v < \sqrt{2G(m_{\rm BH,part,1}+m_{\rm BH,part,2})/r}$, where $m_{\rm BH,part,1}$ and $m_{\rm BH,part,2}$ are the BH particle masses. When these conditions are met, the less massive BH particle is removed from the computational volume, and both its $M_{\rm BH}$ and $m_{\rm BH,part}$ are added to the more massive BH. \textcolor{black}{Linear momentum is conserved during the merger. In the hybrid feedback model, the BH merger prescription further updates the BH spin such that it conserves angular momentum \citep{barausse09}.} Gravitational-wave energy losses are also accounted for by assuming that the merger is of two non-spinning BHs with masses $M_{\rm BH,1}$ and $M_{\rm BH,2}$ and subtracted from the rest mass energy of the descendant for both sub-grid and particle masses \citep{barausse12}.

Below, we explain how gas accretion and AGN feedback are modelled in the fiducial thermal and hybrid simulations. 

\subsubsection{Thermal feedback model (fiducial)}
\label{sssec:colibre-agn-thermal} 

Gas accretion follows the Bondi–Hoyle–Lyttleton (BHL) rate \citep{bondi52, hoyle39}\textcolor{black}{, modified by turbulence \citep[$f_{\rm turb}$ term,][]{ruffert94} and vorticity \citep[$f_{\rm ang}$ term,][]{krumholz05} corrections to account for aspherical inflows, photon trapping, and large-scale gas accretion \citep{izquierdo-villalba23}}:
\begin{equation}
    \dot{M}_{\rm accr}=\dot{M}_{\rm BHL}\frac{f_{\rm turb}\,f_{\rm ang}}{(f^{2}_{\rm turb}+f^{2}_{\rm ang})^{1/2}},
\label{eq:maccr-rate}
\end{equation}
\textcolor{black}{where the BHL rate, $\dot{M}_{\rm BHL}$, is defined as}:
\begin{equation}
    \dot{M}_{\rm BHL}=4\pi G^{2}\frac{\rho_{\rm g}}{c_{\rm s}^{3}}M^{2}_{\rm BH},
\label{eq:maccr-bhl}
\end{equation}
\textcolor{black}{with $G$ the gravitational constant, $\rho_{\rm g}$ the gas density and $c_{\rm s}$ the sound speed of the ambient gas}.

The accretion rate may exceed the Eddington accretion rate ($\dot{M}_{\rm accr}>\dot{M}_{\rm Edd}$), defined as:
\begin{equation}
    \dot{M}_{\rm Edd}=\frac{4\pi Gm_{\rm p}M_{\rm BH}}{\epsilon_{\rm r}\sigma_{\rm T}c},
\label{eq:medd}
\end{equation}
where $m_{\rm p}$ is the proton mass, $\sigma_{\rm T}$ is the Thomson cross-section, $c$ is the speed of light, and $\epsilon_{\rm r}$ is the radiative efficiency \citep{shakura73}. The BH accretion is then capped at $100\times\dot{M}_{\rm Edd}$.

The BH mass grows as:
\begin{equation}
    \dot{M}_{\rm BH}=(1-\epsilon_{\rm r})\,\dot{M}_{\rm accr},
\label{eq:mbh-rate-thermal}
\end{equation}
where $\dot{M}_{\rm accr}$ is the accretion rate onto the BH. The radiative efficiency is fixed at $\epsilon_{\rm r}=0.1$, representing the fraction of accreted mass radiated away as energy. The accretion is assumed to occur equally at accretion-disc and event-horizon scales.

While the BH is accreting at that rate, the associated feedback energy rate is $\dot{E}_{\rm AGN}=\epsilon_{\rm f}\epsilon_{\rm r}\dot{M}_{\rm accr}c^{2}$, where $\epsilon_{\rm f}$ is the fraction of radiative energy that couples to the surrounding gas by heating it or driving a wind through radiation pressure. $\epsilon_{\rm f}$ is calibrated to the $z=0$ BH-to-stellar mass relation. This energy is stored in a reservoir until it exceeds a threshold, at which point it is inserted into the nearest gas particle. This approach ensures that the gas particle is heated to a high temperature, which helps avoid numerical overcooling. Energy injection is purely thermal \citep{booth&schaye09}, representing the effects of radiation (either direct or through radiatively-driven winds that shock on sub-grid scales), with each event delivering:
\begin{equation}
    \Delta E_{\rm AGN,thermal}=\frac{m_{\rm g}k_{\rm B}\Delta T_{\rm AGN}}{(\gamma-1)\mu m_{\rm p}},
\label{eq:deltae-thermal}
\end{equation}
where $\gamma=5/3$ is the adiabatic index, $\mu=0.6$ is the mean molecular weight, and $m_{\rm g}$ is the gas particle mass. $\Delta T_{\rm AGN}$ is the temperature increase applied to the gas particle, and it scales with BH mass as:
\begin{equation}
    \Delta T_{\rm AGN}=10^{9}\,\mathrm{K}\,\left(\frac{M_{\rm BH}}{10^{8}\,\mathrm{M_{\odot}}}\right),
\label{eq:deltat-thermal}
\end{equation}
with $10^{6.5}~{\rm K} < \Delta T_{\rm AGN} < \Delta T_{\rm AGN,max}$, where $\Delta T_{\rm AGN,max}=10^{9.5}\rm\,K$ for the m7 resolution runs and $10^{10}\,\rm K$ for m6 and m5. The dependence of $\Delta T_{\rm AGN}$ on BH mass is an important difference between {\sc COLIBRE} and its predecessor, the {\sc EAGLE} simulations \citep{schaye15}. This implementation enables better sampling of feedback in low-mass galaxies and allows for a gentler early phase of BH growth\footnotemark.

\footnotetext{Heating a single particle generates shock waves that propagate into the surrounding gas and expand due to thermal pressure. Although the feedback is injected \textcolor{black}{into the gas particle closest to the BH,} it typically results in bipolar outflows, as the energy preferentially escapes perpendicular to the disc along the path of least resistance \citep{nobels22}.}

\subsubsection{Hybrid (thermal+jet) feedback model}
\label{sssec:colibre-agn-hybrid} 

The hybrid AGN feedback model \citep{husko25} combines thermal feedback, representing radiation and accretion winds, with kinetic feedback, representing jets. While the underlying BH physics is similar to that in the thermal model, the hybrid implementation introduces key differences in gas accretion and feedback prescriptions. Motivated by both observations and theoretical works, this model incorporates explicit BH spin evolution and accretion disc modelling.

First, BH spin is tracked, evolving through accretion, mergers, jet-induced spin-down, radiation torques, and Lense–Thirring precession. Spin tracking affects both jet orientation and all feedback efficiencies. The BH spin vector is defined as $\vec{a}=\vec{J}_{\rm BH}c/(M_{\rm BH}G)$, where $\vec{J}_{\rm BH}$ is the time-integrated BH angular momentum. Its magnitude $a=|\vec{a}|$ and sign indicate the degree of spin and whether accretion is prograde or retrograde with respect to a sub-grid accretion disc. Seed BHs are initialised with a small spin ($a=0.01$) and random orientation, effectively corresponding to non-rotating BHs. A full description of this model is given in \citet{husko25}.

Second, the accretion disc magnetisation or the magnetic flux around the BH event horizon is modelled, which modulates feedback efficiencies. The model assumes a magnetically arrested disc (MAD) state \citep{narayan03}, where magnetic fields are dynamically important and saturate at a value determined solely by the accretion rate and BH spin \citep{narayan22}.

BH growth follows a similar equation as equation~(\ref{eq:mbh-rate-thermal}), but extended to include the efficiencies of disc winds and jets:
\begin{equation}
    \dot{M}_{\rm BH}=\left(1-\epsilon_{\rm r}-\epsilon_{\rm wind}-\epsilon_{\rm jet}\right)\,\epsilon_{\rm accr}\,\dot{M}_{\rm accr},
\label{eq:mbh-rate-hybrid}
\end{equation}
where $\dot{M}_{\rm accr}$ (equation~\ref{eq:maccr-rate}) here refers to the accretion rate onto a sub-grid accretion disc, rather than directly into the BH event horizon. $\epsilon_{\rm wind}$ and $\epsilon_{\rm jet}$ are the wind and jet efficiencies. $\epsilon_{\rm jet}$ depends on the BH spin and the magnetisation of the disc \citep{tchekhovskoy10}, such that non-spinning BHs ($a=0$) produce no jets ($\epsilon_{\rm jet}=0$). The parameter $\epsilon_{\rm accr}$, equal to 1 in the thermal model, accounts for mass lost to winds between the accretion disc and the BH horizon. \textcolor{black}{BH accretion onto the event horizon is therefore defined as $\dot{M}_{\rm accr,H} = \epsilon_{\rm accr}\dot{M}_{\rm accr}$. For the hybrid AGN feedback model, this quantity is adopted as the BHAR throughout the paper (\S~\ref{sssec:res2-agn-feedback-obs} and \S~\ref{sssec:res2-environment-env})}. \textcolor{black}{The accretion rate onto the sub-grid disc} is also capped at $\dot{M}_{\rm accr}=100\times\dot{M}_{\rm Edd}$. 

The accretion disc operates in three regimes representing different accretion disc properties, determined by \textcolor{black}{$f_{\rm Edd,d}=\dot{M}_{\rm accr}/\dot{M}_{\rm Edd}$ (BH spin-dependent via $\epsilon_{\rm r}$ in the Eddington rate in equation~\ref{eq:medd}). This quantity should not be confused with $f_{\rm Edd}=\dot{M}_{\rm accr,H}/\dot{M}_{\rm Edd}$, defined using the accretion rate at the BH event horizon}. Both thermal and kinetic/jet feedback modes are active in all three regimes, which are:
\begin{itemize}
    \item \textit{Thick disc} (\textcolor{black}{$f_{\rm Edd,d} < 0.01$}): geometrically thick and optically thin. Gas is hot and diffuse, leading to strong advection and negligible radiative efficiency ($\epsilon_{\rm r}\lesssim0.01$). Winds are present, with $\epsilon_{\rm wind}$ depending on spin and magnetisation \citep{sadowski14}. Spinning BHs can also launch strong jets due to magnetic fields. $\epsilon_{\rm accr}$ depends on $f_{\rm Edd,d}$ and the thick-disc radius, following \citet{narayan95} (which contains a free parameter calibrated to the $z=0$ AGN bolometric luminosity function).
    \item \textit{Thin disc} (\textcolor{black}{$0.01 < f_{\rm Edd,d} < 1$}): geometrically thin and optically thick, with a high radiative efficiency $\epsilon_{\rm r}$ determined by BH spin via the innermost stable circular orbit. No disc winds are present ($\epsilon_{\rm wind}=0$). Jets form with power set by the accretion rate and magnetisation state. $\epsilon_{\rm accr}=1$, equivalent to the thermal AGN feedback model.
    \item \textit{Slim disc} (\textcolor{black}{$f_{\rm Edd,d} > 1$}): intermediate state between thick and thin discs. The disc is geometrically thick, but less than the thick disc, as well as advection-dominated and radiatively inefficient ($\epsilon_{\rm r}\lesssim0.01$, dropping as the accretion rate increases). Winds remove most of the inflowing mass, with $\epsilon_{\rm wind}$ set by the accretion disc magnetisation \citep{ricarte23}. Slim discs produce \textcolor{black}{fairly strong} jets. $\epsilon_{\rm accr}=0.01$ is assumed, similar to the thick-disc values, which implies that the upper limit for the net accretion rate can not exceed the Eddington rate: \textcolor{black}{$\mathrm{max}(\dot{M}_{\rm accr,H})=\epsilon_{\rm accr}\,\mathrm{max}(\dot{M}_{\rm accr})=\epsilon_{\rm accr}\,100\times\dot{M}_{\rm Edd}=\dot{M}_{\rm Edd}$}.
\end{itemize}

The total energy available for feedback has a thermal and a kinetic contribution $\dot{E}_{\rm AGN}=\dot{E}_{\rm AGN, thermal}+\dot{E}_{\rm AGN, kinetic}$, stored in separate energy reservoirs. The thermal energy generation and injection follows the fiducial model, but with additional contributions representing both radiation and winds: $\dot{E}_{\rm AGN, thermal}=(\epsilon_{\rm f}\epsilon_{\rm r}+\epsilon_{\rm wind})\epsilon_{\rm accr}\dot{M}_{\rm accr}c^{2}$, where $\epsilon_{\rm f}=0$ for thick and slim discs, and calibrated for thin discs to reproduce the $z=0$ BH-to-stellar-mass relation. 

The kinetic feedback energy production is given by $\dot{E}_{\rm AGN}=\epsilon_{\rm jet}\epsilon_{\rm accr}\dot{M}_{\rm accr}c^{2}$. When the stored kinetic energy is sufficient, the closest particle is chosen on either side of the BH (defined by the two hemispheres along the BH spin vector). These two particles receive opposite velocity kicks, randomly oriented within $7.5^\circ$ of the BH spin axis, releasing per feedback event an amount:
\begin{equation}
    \Delta E_{\rm AGN,jet}=2\times \frac{1}{2}\,m_{\rm g}\,v_{\rm jet}^{2},
\label{eq:deltae-hybrid}
\end{equation}
where the kick velocity per particle $v_{\rm jet}$ is: 
\begin{equation}
    v_{\rm jet}=10^{4.5}\,\mathrm{km\ s^{-1}}\mathrm{}\sqrt{\frac{M_{\rm BH}}{10^{9}\,\mathrm{M_{\odot}}}},
\label{eq:deltav-hybrid}
\end{equation}
with $v_{\rm jet}$ capped at $10^{2.5}\,{\rm km\ s^{-1}}$ and $10^{4.5}\,{\rm km\ s^{-1}}$ at the lower and higher end, respectively, where $10^{4.5}\,{\rm km\ s^{-1}}$ has been selected during the calibration process. The dependence on BH mass improves the sampling of AGN feedback\footnotemark.

\footnotetext{The velocity kick generates shocks that drive jets and inflate hot gas bubbles on Mpc scales \citep[see fig.~1 in][]{husko24}.}

\section{Galaxy properties and sample selection}
\label{sec:prop} 


\textcolor{black}{{\sc COLIBRE} uses the {\sc SOAP} code \citep{mcgibbon25} to compute galaxy properties within apertures centred on the most bound particle. Throughout this series, we adopt a fiducial 50 pkpc spherical aperture, accounting for only particles bound to the subhalo, following \citet{schaye25}. Unless otherwise stated, all galaxy properties are measured within this aperture. As demonstrated in \citetalias{chandro-gomez_inprep}, our results are robust to the exact aperture choice. We summarise below the main galaxy properties used in this work.}

\begin{itemize}
    \item \textit{Stellar mass} ($M_\star$): \textcolor{black}{total mass} of all stellar particles. 
    \item \textit{SFR}: sum of the SFRs of gas particles \textcolor{black}{identified} as star-forming (i.e., locally gravitationally unstable; \citealt{nobels24}). \textcolor{black}{This corresponds to the instantaneous SFR.}
    \item \textit{BH mass} ($M_{\rm BH}$) \textit{and properties}: defined by the most massive BH particle within the fiducial spherical 50 pkpc aperture, considering only particles bound to the subhalo, as used in \citet{husko25}.
    \item \textit{Gas mass in/outflow rate} ($\dot{M}_{\mathrm{in/outflow},fR{\rm 200c}}$): the rate at which gas mass crosses into/out of spherical shells centred at $fR_{\rm 200c}=0.1R_{\rm 200c},\ 0.3R_{\rm 200c}\ \mathrm{or}\ R_{\rm 200c}$, with a shell width of $0.1\times fR_{\rm 200c}$. Here, $R_{\rm 200c}$ is the radius within which the mean density equals 200 times the critical density of the Universe. Gas particles are further classified by temperature: cold ($T_{\rm gas}<10^{3}\,\mathrm{K}$), cool ($10^{3}\,\mathrm{K}<T_{\rm gas}<10^{5}\,\mathrm{K}$), warm ($10^{5}\,\mathrm{K}<T_{\rm gas}<10^{7}\,\mathrm{K}$), and hot ($T_{\rm gas}>10^{7}\,\mathrm{K}$).
    \item \textit{Halo mass} ($M_{\rm halo}$): \textcolor{black}{total mass of the group the particles belong} to, as identified by the 3D FoF algorithm.
\end{itemize}

\subsection{Selecting the high-$z$ MQG sample}
\label{ssec:prop-mq-def} 

\textcolor{black}{We select galaxies according to the following criteria, consistent with common practice:}

\begin{enumerate}
    \item Massive: $\boldsymbol{M_\star>10^{10}\rm M_\odot}$.
    \item \textcolor{black}{Quenched: $\boldsymbol{\mathrm{sSFR}<0.2/t_{\rm age}}$, where the specific star formation rate is defined as $\mathrm{sSFR=SFR}/M_{\star}$ and $t_{\rm age}$ is the age of the Universe at the redshift of interest. Galaxies above this threshold are classified as star-forming. We adopt this redshift- and mass-dependent criterion following \citet{franx08}, as characteristic physical timescales (e.g. dynamical and cooling times) scale with the Hubble time and gas density. In contrast, colour-based selections can be affected by dust-obscured star formation or emission-line contamination \citep{schreiber18}, and may fail to identify quiescent systems that lie outside standard colour cuts \citep{merlin18, baker25b}.}
\end{enumerate}

\textcolor{black}{For the majority of our analysis in \S~\ref{sec:res2}, we will compare the properties of MQGs and massive star-forming galaxies (hereafter MSFGs). The distributions of stellar mass for these populations differ markedly, with MSFGs more skewed towards lower masses than MQGs \citep[based on the galaxy quenched fractions from fig.~9 in][]{chaikin25b}. To properly compare between the two populations and to make use of the full statistical samples, we weight the MSFG distribution to match that of MQGs, ensuring a fair comparison. When computing histograms or medians with percentiles for MSFGs, we apply weights of the form:}
\begin{equation}
    w(M_{\star})=\frac{\phi_{\rm MQG}(M_{\star})}{\phi_{\rm MSFG}(M_{\star})},
    \label{eq:w}
\end{equation}
\textcolor{black}{where $\phi_{\rm i}(M_{\star})$ denotes the stellar mass distribution for each galaxy population. Note that this weighting scheme is not applied in \S~\ref{sssec:res2-agn-feedback-ndens} and \S~\ref{sssec:res2-agn-feedback-obs}, where only the MQG population is analysed.}

\subsection{Environmental tracer $\delta_{\rm \sigma_{\rm subh},rel}(<R)$ definition}
\label{ssec:prop-delta-sigma} 

We also aim to characterise the environment of the MQGs on different scales, beyond their halo masses. Several approaches exist for this purpose; however, we follow recent works in the field, which have employed the density contrast to quantify environment \citep{kimmig25, chittenden25, weller25}. We use the subhalo density contrast as:
\begin{equation}
    \delta_{\rm subh}(<R)=\frac{\rho_{\rm subh}(<R)-\bar{\rho}_{\rm subh}}{\bar{\rho}_{\rm subh}},
\end{equation}
where $\rho_{\rm subh}(<R)$ is the subhalo density within a real-space sphere of radius $R$ while $\bar{\rho}_{\rm subh}$ is the mean subhalo density of the Universe. We compute $\rho_{\rm subh}(<R)=\sum_{i\in R}M_{\mathrm{subhalo},i}/V_{R}$, where $M_{\mathrm{subhalo},i}$ is the total subhalo mass (all bound particles) of each neighbouring galaxy whose galaxy centre, as defined by the most bound particle, resides within the sphere of radius $R$. $V_{R}$ is the volume of the sphere. The second variable is computed as $\bar{\rho}_{\rm subh}=M_{\rm bound}/V_{\rm box}$, where $M_{\rm bound}=\sum_{i\in \rm box}M_{\mathrm{subhalo},i}$ is the mass bound to any galaxy in terms of subhalo hierarchy in the whole simulation box and $V_{\rm box}$ the simulation volume. \textcolor{black}{All subhalos identified by the finder {\sc HBT-HERONS} are included in these definitions, without imposing any mass cut.}

The density contrast characterises the environment of each galaxy. To compare the environments of MQGs and MSFGs, we also examine the distributions of their subhalo density contrasts \citep{kimmig25}. For this purpose, we use the deviation from the mean subhalo density contrast:
\begin{equation}
    \delta_{\rm \sigma_{\rm subh},rel}(<R)=\frac{\delta_{\rm subh}(<R)-\bar{\delta}_{\rm subh}(<R)}{\sigma_{\delta_{\rm subh}}(<R)},
\end{equation}
where $\bar{\delta}_{\rm subh}(<R)$ is the mean subhalo density contrast for the distribution of both MQGs and MSFGs, while $\sigma_{\delta_{\rm subh}}(<R)$ is the standard deviation of that distribution. This metric allows us to quantify the relative subhalo density contrast of each galaxy with respect to the overall population of massive galaxies. We choose three different apertures $R=0.3,\,1.0,\,3.0\,\mathrm{cMpc}$ to capture the environment in a range of local to larger scales in the cosmic web.




\section{The journey of MQGs}
\label{sec:res2} 

We \textcolor{black}{focus on} developing a deeper understanding of the physical drivers that shape the properties of MQGs in the {\sc COLIBRE} model. In \S~\ref{ssec:res2-agn-feedback}, we investigate what suppresses star formation so early in time for such massive systems. We focus on the role of AGN feedback because it is most likely to be the dominant mechanism driving quiescence\footnotemark. In \S~\ref{ssec:res2-environment}, we explore the link between quenching and a galaxy's environment, and how this shapes its subsequent evolution. Finally, in \S~\ref{ssec:res2-rejuvenation}, we examine the long-term fate of MQGs and their potential to undergo rejuvenation episodes. 

\footnotetext{SN feedback is not expected to quench galaxies in this massive regime \citep{benson03}, and the quenching timescales analysed in \textcolor{black}{sec.~4.3.2 of \citetalias{chandro-gomez_inprep}} (with median values up to $t_{\rm q}\approx 0.6\,\mathrm{Gyr}$) make this even less plausible.} 

For this part of the analysis, we fix the selection redshift for our samples at $z=3$, which provides sufficient statistics, and we have verified that the results are similar when the sample is selected at another redshift $z\ge2$. \textcolor{black}{We compare MQGs and MSFGs using the weighting scheme described in \S~\ref{ssec:prop-mq-def} to match their stellar mass distributions at $z=3$. We then} follow the merger histories of both MQGs and MSFGs, tracking their evolution backwards and forwards in time relative to the selection redshift, using {\sc HBT-HERONS}, \textcolor{black}{noting that their distributions naturally diverge at $z\ne3$}. 
We assess the effects of AGN feedback in the fiducial L200m6 simulation, which includes only a thermal model and contains 470 $z=3$ MQGs (following the selection in \S~\ref{ssec:prop-mq-def}); and in the L200m7h simulation, which includes the hybrid (thermal+jet) model and contains 310 $z=3$ MQGs. 
The results from these two simulations should be compared qualitatively, given their different mass resolutions and free parameter values in the calibration.

\subsection{AGN feedback (selection time or ``present'')}
\label{ssec:res2-agn-feedback} 

As a first step toward understanding the role of AGN feedback in MQGs, we calculate the cumulative energy injected by the most massive BH particle within the fiducial aperture ($50\,\rm pkpc$). \textcolor{black}{We verified that this most massive BH corresponds to the BH associated with the galaxy of interest, despite the aperture being much larger than the typical size of these galaxies (see fig.~8 of \citetalias{chandro-gomez_inprep})}. The value $E_{\rm inj,BH}=\sum_{\rm event}{\Delta E_{\rm AGN}}$ is the cumulative sum over all the AGN feedback events associated with that BH particle. In the thermal simulation, this corresponds to the thermal energy deposited into gas particles within the BH's smoothing kernel (accumulating the contribution from equation~ \ref{eq:deltae-thermal}); in the hybrid simulation, this corresponds to thermal energy (as in the fiducial simulation) as well as kinetic energy from jets (accumulating the contributions from equations~\ref{eq:deltae-thermal}~and~\ref{eq:deltae-hybrid}). Note that this does not include energy contributions from BHs that previously merged with this BH particle.

\begin{figure}
\centering
\includegraphics[trim={0 0 0 0},clip, width=0.48\textwidth]{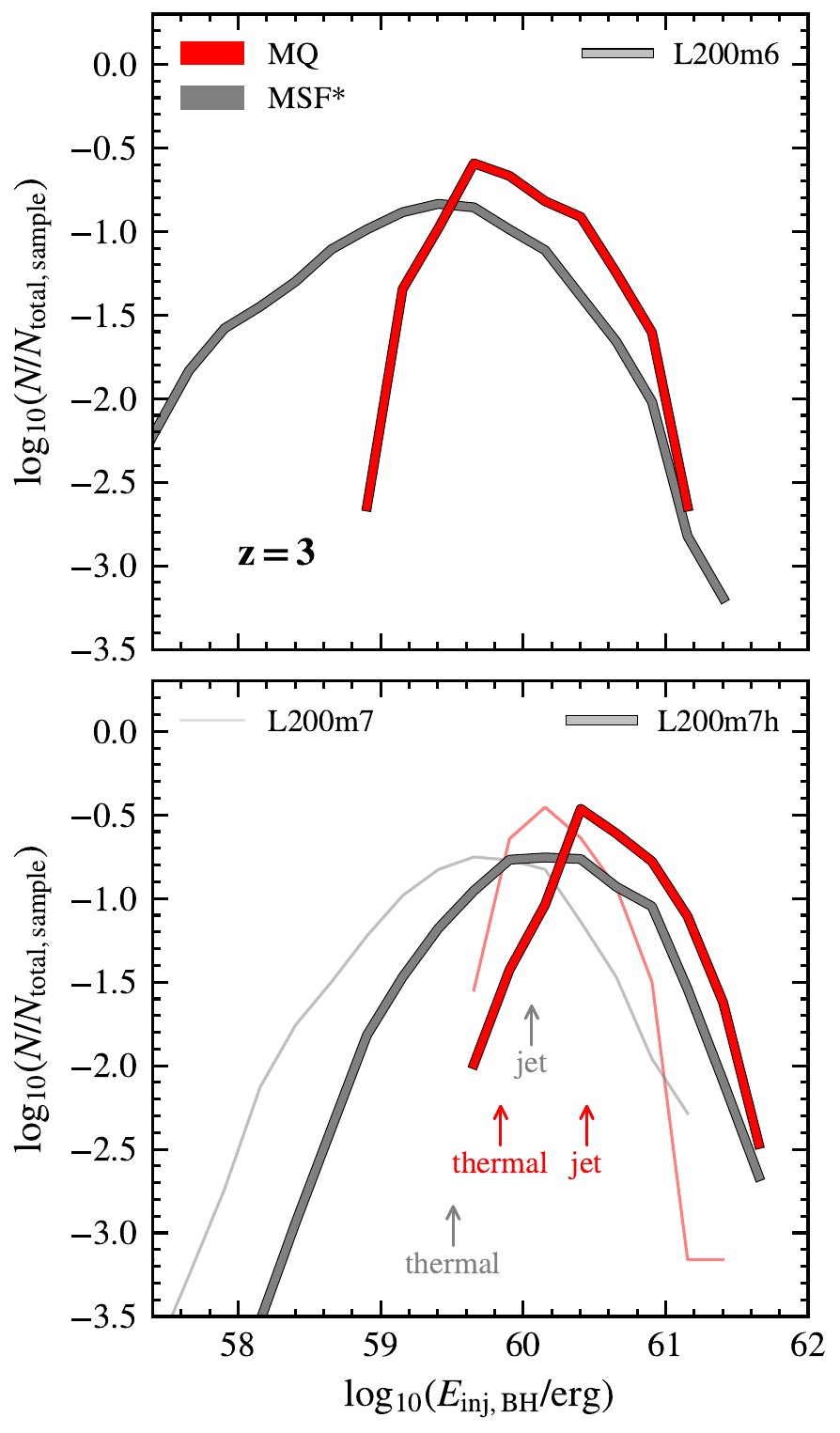}
\caption{Normalised histogram \textcolor{black}{(by the total number of galaxies in each sample)} of cumulative BH-injected energy for MQGs (red) and \textcolor{black}{the weighted} MSFGs (grey) at $z=3$ in the fiducial L200m6 simulation (\textit{top panel}) and the hybrid L200m7h simulation (\textit{bottom panel}). Vertical arrows in the bottom panel mark the median contributions from the jet and thermal components in the hybrid case. Narrow solid lines in the background show the L200m7 results, included to illustrate the AGN feedback model effect.}
\label{fig:bh-energy} 
\end{figure}

Fig.~\ref{fig:bh-energy} shows the normalised histograms of the cumulative BH energy for MQGs (solid red curves) and \textcolor{black}{the weighted} MSFGs (solid grey curves) at $z=3$. The top panel presents results for the thermal L200m6 simulation, while the bottom shows the hybrid L200m7h simulation. In both cases, the MQGs and MSFGs peak at different values: BHs in star-forming systems inject, on average, less energy. In the thermal simulation, the median cumulative energy is $\approx10^{59.8}\,\rm erg$ for MQGs compared to $\approx10^{\textcolor{black}{59.3}}\,\rm erg$ for MSFGs, while in the hybrid simulation $\approx10^{60.5}\,\rm erg$ and $\approx10^{\textcolor{black}{60.2}}\,\rm erg$, respectively. These differences \textcolor{black}{point to AGN feedback as the driver of quenching}.

The hybrid case broadly resembles the thermal case, but the total (thermal+jet) energy is larger. 
To assess the effect of the AGN feedback model independently of volume and resolution, we include the L200m7 simulation, which shows that the hybrid simulation still injects more feedback energy compared to the thermal AGN feedback run. This may appear contradictory since quenching is less efficient in the hybrid simulation (see Fig.~\ref{fig:ndens-boxes} \textcolor{black}{or fig.~2 of \citetalias{chandro-gomez_inprep}}). Note that there is more energy injection overall in the m7 resolution runs, though.

The cumulative BH-injected energy in the hybrid case consists of a combination of thermal plus kinetic jet energy ($E_{\rm inj,BH}=E_{\rm thermal}+E_{\rm jet}$). Vertical arrows mark the median thermal, $\mathrm{med}(E_{\rm thermal})$, and jet, $\mathrm{med}(E_{\rm jet})$, contributions, showing that the jet model dominates in both MQGs and MSFGs, while the thermal contribution falls below that of L200m7. This suggests that jets require longer timescales to quench galaxies.

\begin{figure*}
\centering
\includegraphics[width=\textwidth]{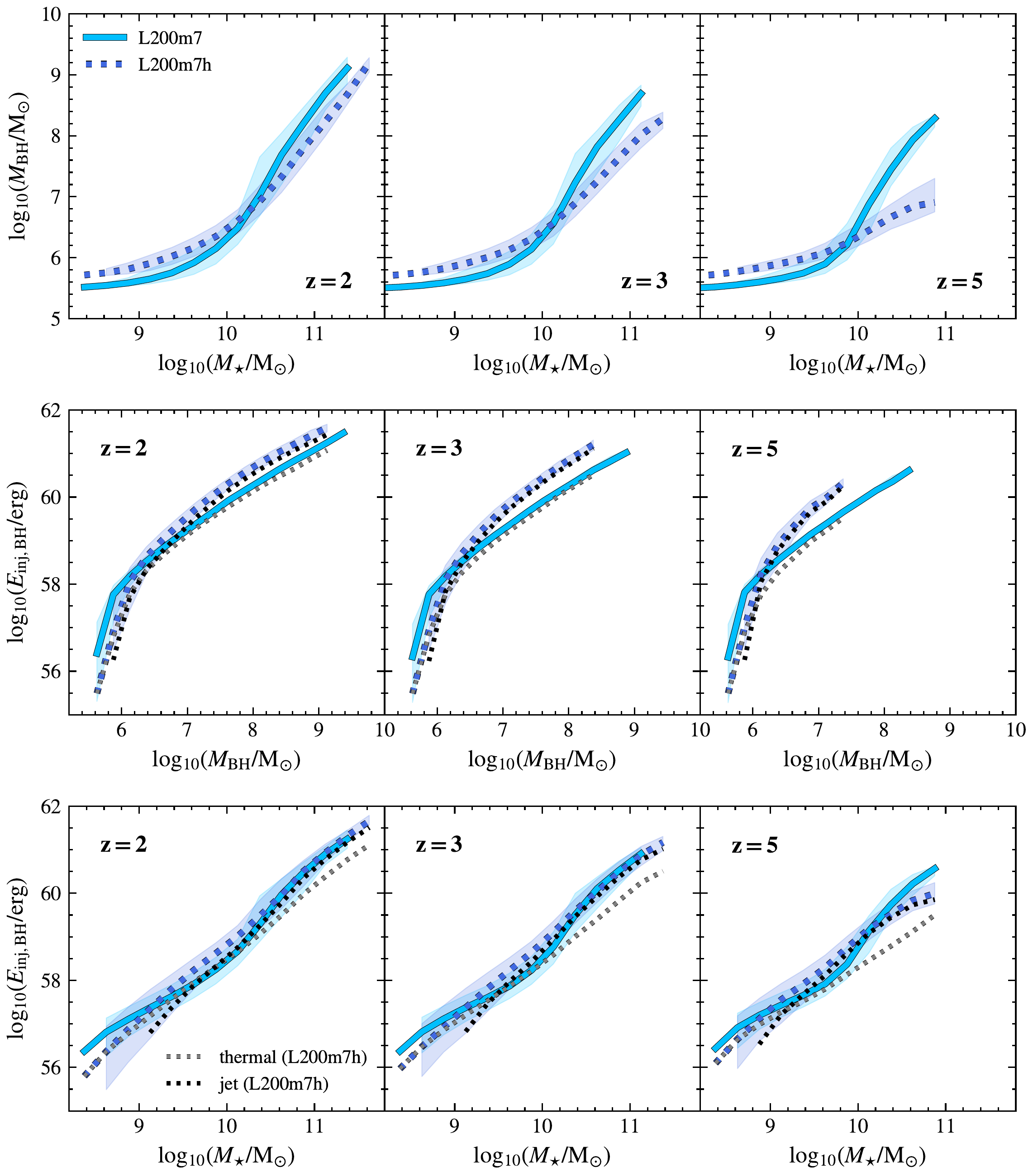}
\caption{BH mass-stellar mass relation (\textit{first row}), BH-injected energy-BH mass relation (\textit{second row}) and BH-injected energy-stellar mass relation (\textit{third row}) at different redshifts ($z=2, 3, 5$; columns). Lines show median values with the 16th and 84th percentile ranges as shaded regions: solid light blue for L200m7 and dotted dark blue for L200m7h. Grey and black narrow dotted lines indicate the thermal and jet model contributions, respectively, in L200m7h.}
\label{fig:app-thermal-hybrid} 
\end{figure*}
\begin{figure*}
\centering
\includegraphics[width=\textwidth]{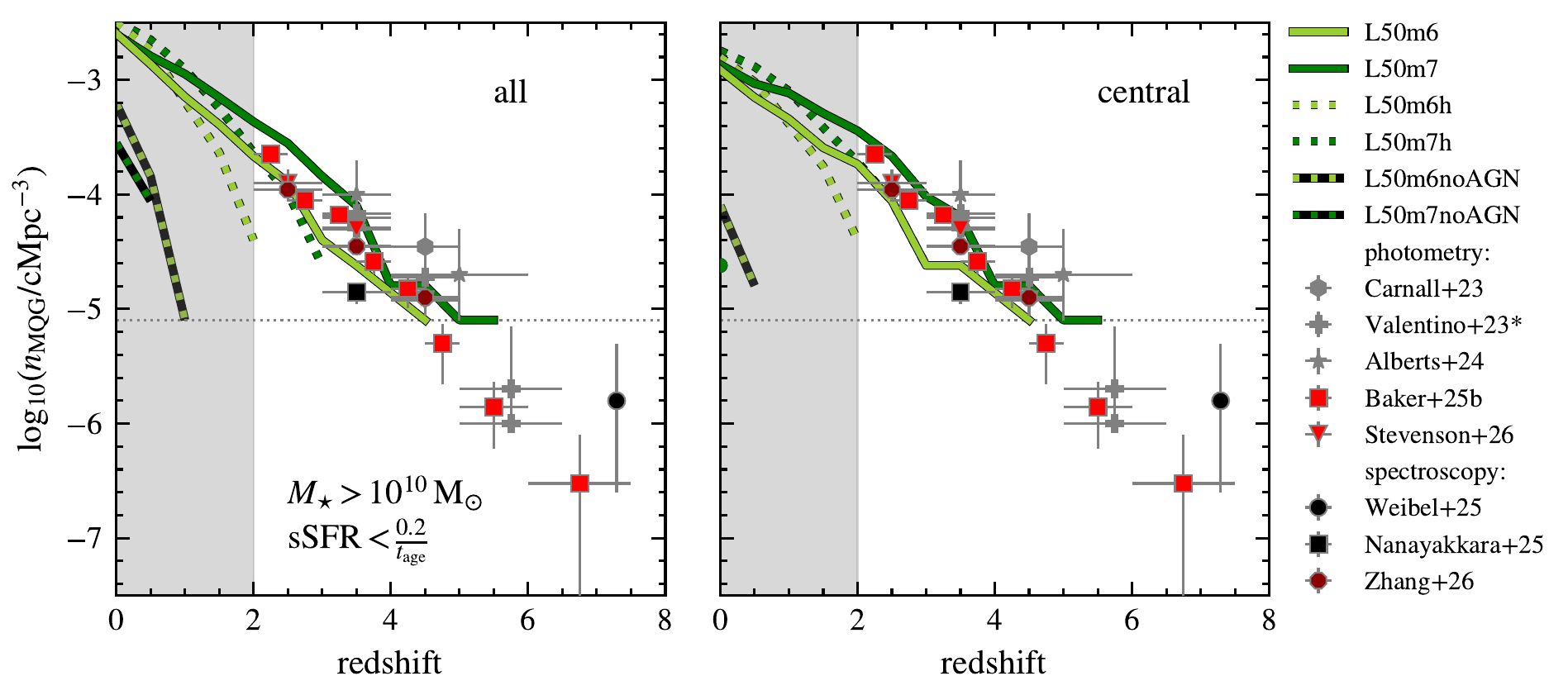}
\caption{Comoving number density of MQGs, defined by $M_{\star}>10^{10}\,\rm \textcolor{black}{M_\odot}$ and $\mathrm{sSFR}<0.2/t_{\rm age}$, as a function of redshift. Lines represent predictions from various \textcolor{black}{$L=50\,\mathrm{cMpc}$} {\sc COLIBRE} boxes (spanning different volumes, resolutions, and AGN feedback models) in Table~\ref{tab:runs}, with the selection at $0 \le z \le 8$ \textcolor{black}{(the region of interest, $z\ge2$, is not shaded)}. \textcolor{black}{Lines that terminate at $z<8$ indicate that no galaxies are identified as MQGs beyond those redshifts}. These are compared with the latest \textit{JWST} observational estimates: black (smaller area) and dark red (larger area) points show spectroscopic measurements, while grey (smaller area) and red (larger area) points correspond to photometric measurements. Observational data are taken from \citet{carnall23, valentino23, alberts24, baker25, weibel25, nanayakkara25, stevenson25, zhang25}. \textcolor{black}{The horizontal dotted line marks the threshold corresponding to one galaxy in the simulation volume. \textit{Left panel}: all (central plus satellite) MQGs from {\sc COLIBRE}. \textit{Right panel}: only central MQGs from {\sc COLIBRE}.}}
\label{fig:ndens-boxes} 
\end{figure*}

We can \textcolor{black}{further} examine how the differences between thermal and hybrid AGN feedback models \textcolor{black}{(in this paper or in the number densities of MQGs in fig.~2 of \citetalias{chandro-gomez_inprep})} arise from BH-related physics \textcolor{black}{by comparing L200m7 and L200m7h}.
Fig.~\ref{fig:app-thermal-hybrid} shows the median relations (with the 16th and 84th percentile ranges) for all galaxies in the simulated volume: L200m7 in dotted dark blue and L200m7h in solid light blue, at $z=2$ (first column), $z=3$ (second), and $z=5$ (third). \textcolor{black}{The thermal and jet contributions for the hybrid model are shown by thin grey and black dotted lines, respectively}. 

The top row shows the BH–stellar mass relation. At high redshift ($z=5$), BHs in the hybrid model grow less efficiently, especially in \textcolor{black}{massive} galaxies ($M_{\star}>10^{10}\,\mathrm{M_{\odot}}$), which are the focus of this study.
The middle row presents BH-injected energy as a function of BH mass. Although BHs are less massive in the hybrid model at early times, they are more efficient at injecting energy, with the jet model dominating the contribution, consistent with \textcolor{black}{Fig.~\ref{fig:bh-energy} and} fig.~14 in \citet{husko25}. This is an indication of BH self-regulating its evolution \citep{booth&schaye09}. 
The bottom row shows BH-injected energy versus stellar mass. At $z=5$, the most massive galaxies in the hybrid simulation receive less injected energy than in the thermal case. By $z\approx2-3$, however, the hybrid simulation reaches comparable or even higher values. Despite this, the thermal-only model still produces slightly more quenching (see fig.~2 of \citetalias{chandro-gomez_inprep}). Examining the jet and thermal contributions in the hybrid simulation (thin lines) suggests that jets act on longer timescales to effectively quench galaxies, as the thermal component remains consistently below the thermal-only case. \textcolor{black}{This is a key result: quenching depends not only on the total amount of injected energy, but also on how that energy is injected and couples to the surrounding gas.}

\begin{table}
	\centering
	\caption{Energy contribution from each BH accretion disc state (thick, thin, or slim) or AGN feedback model (thermal or jet) in the hybrid L200m7h simulation \textcolor{black}{for MQGs and the weighted MSFGs at $z=3$}. $\mathrm{med}(E_{i}/E_{\rm inj,BH})$ denotes the median ratio of each cumulative component $E_{i}$ relative to the total BH-injected energy, $E_{\rm inj,BH}=\sum_{i} E_{i}$.}
	\label{tab:hybrid-energy}
	\begin{tabular}{ccc}
        \hline
        property at $z=3$ & \multirow{2}{*}{MQ} & \multirow{2}{*}{MSF} \\
        (L200m7h) & & \\
        \hline
        $\mathrm{med}(E_{\rm thick,thermal}/E_{\rm inj,BH})$ & 0.00 & 0.00 \\
        $\mathrm{med}(E_{\rm thin,thermal}/E_{\rm inj,BH})$ & 0.16 & 0.19 \\
        $\mathrm{med}(E_{\rm slim,thermal}/E_{\rm inj,BH})$ & 0.04 & 0.04 \\
        $\mathrm{med}(E_{\rm thick,jet}/E_{\rm inj,BH})$ & 0.00 & 0.00 \\
        $\mathrm{med}(E_{\rm thin,jet}/E_{\rm inj,BH})$ & 0.79 & 0.76 \\
        $\mathrm{med}(E_{\rm slim,jet}/E_{\rm inj,BH})$ & 0.00 & 0.04 \\
        \hline
        
	\end{tabular}
\end{table}

Table~\ref{tab:hybrid-energy} shows the median energy contribution from each feedback model \textcolor{black}{for MQGs and the weighted MSFGs at $z=3$}, split by accretion disc state (thick, thin, slim; see \S~\ref{sssec:colibre-agn-hybrid}). In all massive galaxies, the thin disc dominates, contributing over 90\% of the injected energy. This is consistent with \citet{husko25}, who noted that the thin disc behaves similarly to the fiducial thermal model when $f_{\rm Edd}<0.3$ (thermal feedback is then dominant, since $\epsilon_{\rm acc}=1$; see their table~1). However, in the regime $0.3<f_{\rm Edd}<1$, jets dominate in the thin disc. In MQGs, 79\% of the cumulative energy comes from the thin-disc jet model, compared to \textcolor{black}{76}\% in MSFGs. The thin-disc thermal contribution is also non-negligible. Importantly, in $\approx95\%$ of MQGs, the thin-disc jet model dominates the BH energy budget, peaking at $z\approx3$ \citep{husko25}. We find that $\approx86\%$ of the quenched systems are still in that state at $z=3$.

\textcolor{black}{We also find a non-negligible ($\approx23\%$ and $\approx24\%$) population of active galaxies whose supermassive BHs inject large amounts of energy (exceeding the MQG median values: $\gtrsim10^{59.8}\,\rm erg$ for L200m6, and $\gtrsim10^{60.5}\,\rm erg$ for L200m7h, respectively) yet remain star-forming at $z=3$. Tracking their descendants shows that in L200m6 (L200m7h), 26\% (58\%) quench by $z=2.5$ and 35\% (71\%) by $z=2$, excluding $\approx5\%$ (2\%) that disappear through mergers by $z=2$ (they are not the main progenitor). There is a small fraction, 7\% (5\%), that were already quenched at $z=3.125$ but experience a rejuvenation phase (more details about rejuvenation in \S~\ref{ssec:res2-rejuvenation}). These fractions are higher than for MSFGs with lower BH energy injection (lower than the MQG median values), for which 12\% (15\%) quench by $z=2.5$ and 26\% (39\%) by $z=2$, with 9\% (7\%) merging into other systems. Only 2\% (2\%) of these lower-BH-energy systems were quenched at $z=3.125$. Note that all quoted fractions account for the stellar mass weighting scheme applied to match MQG and MSFGs (equation~\ref{eq:w}).}

\textcolor{black}{Some of these high-BH-energy systems are therefore likely on the verge of quiescence or have undergone temporary rejuvenation. They also tend to have sSFR values near the imposed threshold, with a median $\mathrm{log}_{10}(\mathrm{sSFR/yr^{-1}})\approx-9.5$ in L200m6 (compared to the threshold of $\approx-10$), whereas lower-BH-energy MSFGs have higher values ($\approx-9.0$), corresponding to higher $\rm H_2$ and dust fractions. Nevertheless, a significant fraction of the high-BH-energy galaxies do not fall into any of these categories. More isotropic gas inflows, which may reduce the efficiency of AGN feedback, or other processes could also play a role, although investigating this further is beyond the scope of this paper.}

\textcolor{black}{Overall, these results highlight the central role of AGN feedback and motivate a more detailed investigation of its impact.}

\subsubsection{Number densities of MQGs in non-AGN feedback runs}
\label{sssec:res2-agn-feedback-ndens}

\textcolor{black}{The most direct way to isolate the impact of AGN feedback on these galaxies is to use {\sc COLIBRE} runs that do not include any AGN feedback model (neither thermal nor hybrid). In these simulations, BHs are not directly seeded (i.e. the prescriptions in \S~\ref{ssec:colibre-agn} are not applied). We use small {\sc COLIBRE} boxes with $L=50\,\mathrm{cMpc}$ per side, whose features are presented in Table~\ref{tab:runs}. For this box size, two resolution levels (m6 and m7) are available, allowing us to test the impact of AGN feedback across three configurations: thermal AGN feedback, hybrid AGN feedback, and no AGN feedback; and two resolutions. In the non-AGN runs, AGN feedback is simply switched off, with all other free parameters kept identical to those in the thermal AGN feedback runs. Although the box size is not large enough for a statistical study of MQGs, it is sufficient to assess the role of AGN feedback.}

\textcolor{black}{We compute the most direct observable related to the existence of MQGs, namely the evolution of their number density, following the approach presented in \citetalias{chandro-gomez_inprep} (sec.~4.1). In this case, we compare predictions from the non-AGN runs with those from simulations that include AGN feedback. The resulting number densities are shown in Fig.~\ref{fig:ndens-boxes}, using the selection defined in \S~\ref{ssec:prop-mq-def}, computed at intervals of 0.5 in redshift over $z=0-8$, together with JWST observational measurements. The redshift range is extended down to $z=0$ for completeness, although the main focus of this series of papers is $z\ge2$. When no galaxies are identified as MQGs at a given redshift, the corresponding redshift bin is omitted, causing some lines to terminate at $z<8$.}

\textcolor{black}{In the left panel of Fig.~\ref{fig:ndens-boxes}, which shows the results for all MQGs irrespective of whether they are centrals or satellites, we find that the runs without AGN feedback (shown as black-green dashed lines) produce significantly lower number densities, with no MQGs at $z>1$. This provides clear evidence that AGN feedback is the primary quenching mechanism for the galaxies of interest in this series of papers. Furthermore, as shown in the right panel, which presents the same results but considering only central galaxies, the number densities in the runs without AGN feedback are much lower than in the left panel, unlike the runs with AGN feedback, where the MQG population is dominated by centrals. This indicates that, in the non-AGN runs, most systems that quench at $z\leq1$ are satellites, suggesting that environmental effects dominate quenching in these runs. This interpretation is supported by recent work showing that environmental quenching is only efficient at $z>3$ for low-mass systems with $M_{\star}<10^{10}\,\mathrm{M}_{\odot}$ \citep{doven26}.}

\textcolor{black}{A similar argument could be made for the hybrid runs (dotted lines); however, the differences between thermal and hybrid models arise because BHs grow more slowly in the hybrid mode and jet feedback requires longer timescales to affect the galaxy (\S~\ref{ssec:res2-agn-feedback}). We also observe the effects of cosmic variance due to the smaller simulation volumes, as well as resolution differences between the m6 and m7 runs, both discussed in detail in \citetalias{chandro-gomez_inprep} (sec.~4.1).}

\textcolor{black}{These results demonstrate that, within the {\sc COLIBRE} model, AGN feedback is required to quench massive galaxies in the early Universe. 
\citet{chaikin26} provide further support for this interpretation, showing the importance of super-Eddington accretion in {\sc COLIBRE}. Using $L=100\,\mathrm{cMpc}$ boxes with fiducial thermal AGN feedback and different limits on BHARs (sub-Eddington, Eddington and super-Eddington), they find that super-Eddington accretion enables the formation of MQGs at very high redshift ($z\gtrsim6$). This distinguishes {\sc COLIBRE} from other current galaxy formation models, as it allows for more rapid BH growth, partly due to the explicit modelling of the cold gas phase. Although super-Eddington events are rare, they dominate BH growth in massive galaxies at these high redshifts, as shown in fig.~4 of \citet{chaikin26} and in \citet{husko25b}.}

\subsubsection{AGN properties of MQGs: comparison with observations}
\label{sssec:res2-agn-feedback-obs}

\begin{figure*}
\centering
\includegraphics[trim={0 0 0 0},clip, width=\textwidth]{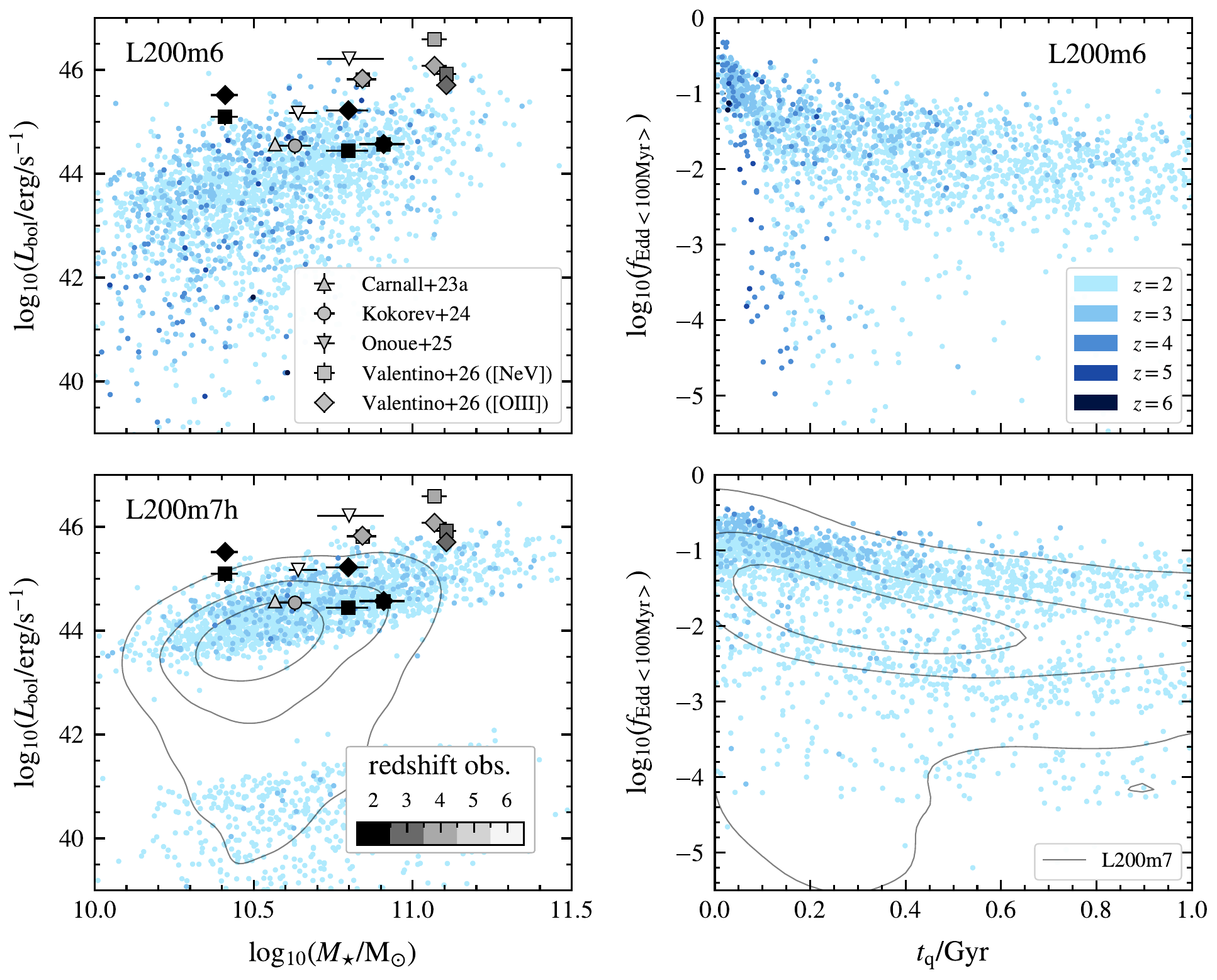}
\caption{\textcolor{black}{Scatter plots for individual MQGs, with points colour-coded by selection redshift, as indicated in the legend in the top-right panel, for the L200m6 (\textit{top}) and L200m7h (\textit{bottom}) simulations. Contours in the bottom panels show the 30\%, 60\% and 90\% distributions of all $z\ge2$ MQGs in the L200m7 simulation. \textit{Left panels}: instantaneous AGN bolometric luminosity of the most massive BH bound to the galaxy (\textcolor{black}{$L_{\rm bol}=\epsilon_{\rm r}\dot{M}_{\rm accr}c^{2}$ for the thermal AGN feedback model, and $L_{\rm bol}=\epsilon_{\rm r}\epsilon_{\rm accr}\dot{M}_{\rm accr}c^{2}$ for the hybrid AGN feedback model}) as a function of stellar mass. Observational data from \citet{carnall23b, kokorev24, onoue25, valentino_inprep}; are overplotted, with redshift indicated by the inset colour bar. \textit{Right panels}: Eddington ratio of the most massive BH bound to the galaxy (\textcolor{black}{$f_{\rm Edd}=\dot{M}_{\rm accr}/\dot{M}_{\rm Edd}$ for the thermal model, and $f_{\rm Edd}=\epsilon_{\rm accr}\dot{M}_{\rm accr}/\dot{M}_{\rm Edd}$ for the hybrid model}), averaged over $100\,\mathrm{Myr}$, as a function of quenching timescale (defined in \citetalias{chandro-gomez_inprep}). \textcolor{black}{Note that this quantity is averaged over $\min(100\,\mathrm{Myr}, \Delta t_{\rm out})$, where $\Delta t_{\rm out}$ is the time interval between consecutive simulation outputs.}}}
\label{fig:agn-props} 
\end{figure*}

\textcolor{black}{After establishing the crucial role of AGN feedback in quenching these systems, we investigate several AGN-related properties. Fig.~\ref{fig:agn-props} shows predictions for MQGs in L200m6 (top panels) and L200m7h (bottom panels). The left panels show the instantaneous AGN bolometric luminosity as a function of stellar mass. The AGN bolometric luminosity is defined as \textcolor{black}{$L_{\rm bol}=\epsilon_{\rm r}\dot{M}_{\rm accr}c^{2}$ for the thermal AGN feedback model (from equation~\ref{eq:mbh-rate-thermal}), with a fixed radiative efficiency of $\epsilon_{\rm r}=0.1$. For the hybrid AGN feedback model (from equation~\ref{eq:mbh-rate-hybrid}), $L_{\rm bol}=\epsilon_{\rm r}\dot{M}_{\rm accr,H}c^{2}=\epsilon_{\rm r}\epsilon_{\rm accr}\dot{M}_{\rm accr}c^{2}$, where $\epsilon_{\rm r}$ depends on BH spin and accretion mode, while the accretion efficiency, $\epsilon_{\rm accr}$, depends on the accretion mode as well and, in the thick-disc regime, on the sound speed of the ambient gas \citep[see][]{husko25}}. The right panels present the $100\,\rm Myr$-averaged Eddington ratio \textcolor{black}{as a function of quenching time, $t_{\rm q}$. For the thermal model, $f_{\rm Edd}=\dot{M}_{\rm accr}/\dot{M}_{\rm Edd}$ (combining equations~\ref{eq:maccr-rate}~and~\ref{eq:medd}), while for the hybrid model $f_{\rm Edd}=\dot{M}_{\rm accr,H}/\dot{M}_{\rm Edd}=\epsilon_{\rm accr}\dot{M}_{\rm accr}/\dot{M}_{\rm Edd}$. Note that when the time interval between consecutive simulation outputs, $\Delta t_{\rm out}$, is shorter than $100\,\rm Myr$, the quantity is averaged over $\min(100\,\mathrm{Myr}, \Delta t_{\rm out})$. The quenching time is defined in sec.~3.2 of \citetalias{chandro-gomez_inprep} as the lookback time from the selection redshift, not from $z=0$, when the SFR last dropped below 10\% of its peak value.} 
The scatter points correspond to individual MQGs at $z\ge2$, colour-coded by selection redshift. To isolate the effect of the AGN feedback model from volume and resolution, we include the 30\%, 60\% and 90\% distribution contours for all MQGs identified in the L200m7 simulation in the bottom panels.}

\textcolor{black}{In the left panels, more massive systems tend to host more luminous AGN, although with substantial scatter. MQGs in L200m7h reach larger stellar masses \textcolor{black}{and exhibit a gap in luminosity around $L_{\rm bol}\approx10^{41}$--$10^{43}\,\mathrm{erg\,s^{-1}}$, reflecting the transition between different BH accretion disc regimes (\S~\ref{sssec:colibre-agn-hybrid}). This gas is driven primarily by differences in the accretion efficiency, $\epsilon_{\mathrm{acc}}$ \citep[see fig.~4 of][]{husko25}, and to a lesser extent by variations in the radiative efficiency, $\epsilon_{\mathrm{r}}$. The upper-luminosity sequence corresponds to the thin and slim disc regimes, while the lower-luminosity sequence corresponds to the thick disc.}
Although the trend is clearer for the $100\,\rm Myr$ averaging than for the instantaneous values, due to the stochastic nature of BH accretion, we plot only the instantaneous values, since observational estimates trace AGN activity on timescales typically shorter than $1\,\mathrm{Myr}$ \citep[e.g.][]{schawinski15}.}

\textcolor{black}{We compare these predictions with recent spectroscopic observations of quasars (i.e. luminous AGN with $L_{\rm bol}\gtrsim10^{45}\mathrm{erg\,s^{-1}}$) hosted by MQGs. These quasar-like MQGs are not representative of the overall population, instead comprising a small fraction of MQGs with strong emission lines \citep[e.g. less than $10\%$ in][]{valentino_inprep}. Moreover, these systems lie in the low-luminosity tail of the quasar population \citep{he24}. Most MQGs instead host lower-luminosity AGN \citep[e.g.][]{bugiani25} or have no AGN detected \citep[e.g. non-detections in][]{ito22, stevenson25}. }

\textcolor{black}{Observationally, stellar masses are typically derived from SED fitting.
On the other hand, bolometric luminosities are inferred either from the UV/optical continuum luminosity for unobscured sources \citep[e.g. the continuum at $5100\,\text{\AA}$ for the two $z>6$ galaxies in][]{onoue25}, or from emission-line luminosities in obscured systems, where the ionised gas is powered by AGN accretion. For example, \citet{kokorev24} and \citet{carnall23b} use the broad H$\alpha$ line, while the \citet{valentino_inprep} sample uses the [OIII] and [NeV] emission lines (6 galaxies with [OIII] and [NeV] coverage and detected broad H$\alpha$). In any case, empirical conversions are used to relate the continuum/emission-line luminosity to the bolometric luminosity. The fiducial L200m6 simulation is broadly consistent with most of these observations, although it does not reproduce the most extreme systems reported by \citet{onoue25} at $z>6$ (where simulation statistics are limited), nor some of the most luminous sources in \citet{valentino_inprep} at $z\approx4$. However, we do not include all {\sc COLIBRE} outputs in the redshift range of interest, only those at integer redshifts. Overall, the fiducial thermal model provides a slightly better match to the observations than the hybrid one, consistent with the results found in \citetalias{chandro-gomez_inprep}.}

\begin{figure*}
\centering
\begin{subfigure}[b]{0.46\textwidth}
   \includegraphics[trim={0 0 0 0},clip, width=\textwidth]{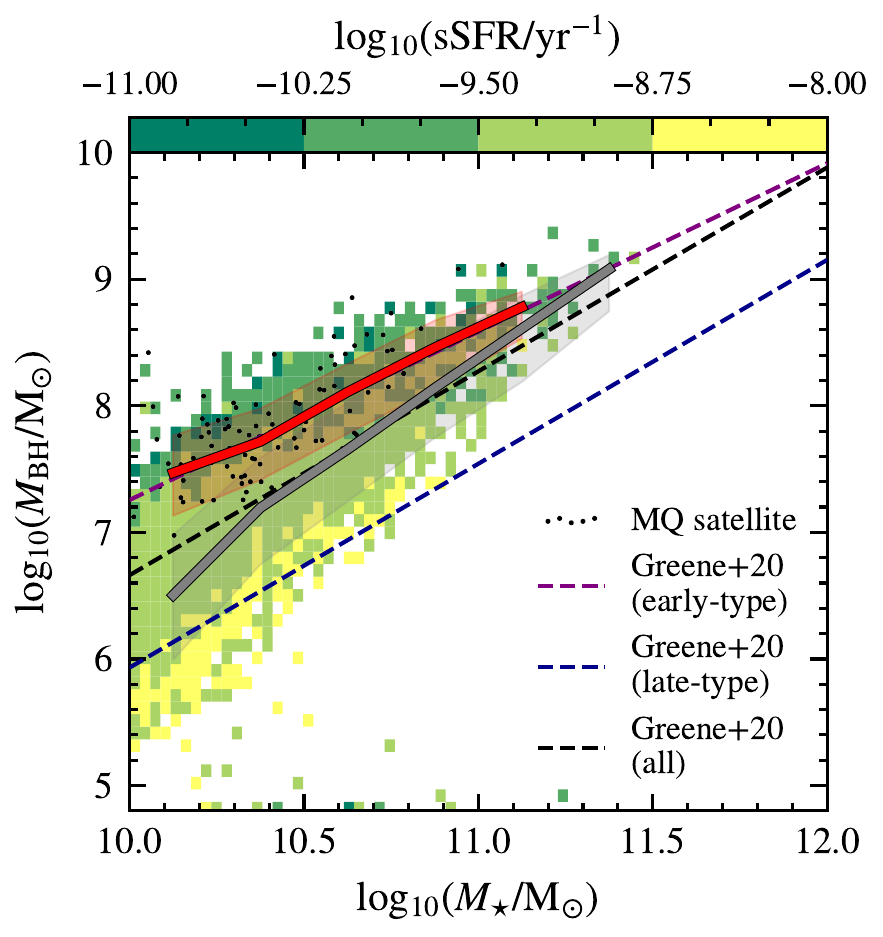}
\end{subfigure}
\begin{subfigure}[b]{0.49\textwidth}
   \includegraphics[trim={0 0 0 0},clip, width=\textwidth]{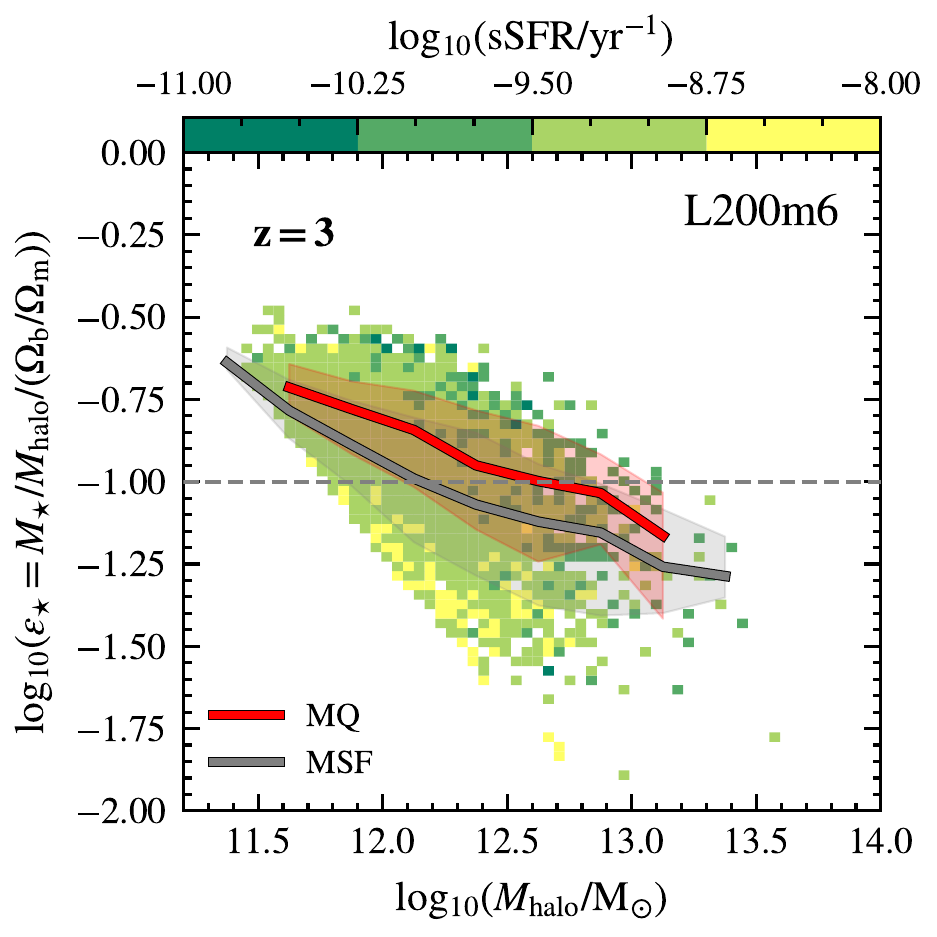}
\end{subfigure}
\caption{BH mass-stellar mass relation for all massive systems (\textit{left panel}) and star formation efficiency, defined as $\varepsilon_{\star}=M_{\star}/M_{\rm halo}/(\Omega_{\rm b}/\Omega_{\rm m})$, versus halo mass (\textit{right panel}) for central massive systems selected at $z=3$ in L200m6. \textcolor{black}{Background 2D maps show the distribution of massive galaxies ($M_{\star}>10^{10}\,\rm M_{\odot}$), colour-coded by the median sSFR in each bin.} Solid lines show the median values for MQGs (red) and MSFGs (grey), with shaded regions indicating the 16th and 84th percentile range. Satellite MQGs are marked with black dots in the left panel; \textcolor{black}{while observational local scaling relations for \textcolor{black}{early- and late-type galaxies as well as} for the full galaxy population from \citet{greene20} are shown as dashed \textcolor{black}{purple, dark blue and black} lines, respectively}. In the right panel, the dashed horizontal line marks a reference value of 10\% star formation efficiency.}
\label{fig:mbh-mstars-eff-thermal} 
\end{figure*}

\textcolor{black}{In the right panels, MQGs that quenched more recently (shorter $t_{\rm q}$) exhibit higher Eddington ratios, peaking around $t_{\rm q}=0$. This indicates that AGN activity reaches its maximum near the quenching time, when BHAR is strongest. Fig.~\ref{fig:images-h} in Appendix~\ref{appendix:images} illustrates this behaviour for individual MQGs in L200m6, showing peaks in BH accretion immediately before quenching happens. After this phase, the gas surrounding the BH is likely depleted or expelled from the galaxy centre, reducing the fuel for accretion and lowering $f_{\rm Edd}$.}

\textcolor{black}{For galaxies that quenched earlier ($t_{\rm q}>0.1\,\mathrm{Gyr}$), the scatter increases significantly, reflecting a range of reaccretion histories and accretion rates. Some systems may begin to reaccrete gas, while others remain depleted, leading to a broad distribution of Eddington ratios (as also illustrated in Appendix~\ref{appendix:images}). For example, in L200m6, around 39\% of the $z=3$ MQGs reach a minimum in the instantaneous BHAR after quenching and later experience a renewed accretion episode before the selection time, where the BHAR increases by more than 0.5 dex relative to that minimum value. 
The median reignition timescale, defined as the time between quenching and the onset of significant reaccretion, is $\approx200\,\mathrm{Myr}$. This interpretation is supported by the tail of systems with very low $f_{\rm Edd}$ at $t_{\mathrm{q}}\lesssim200\,\mathrm{Myr}$, which likely correspond to cases where AGN feedback has efficiently evacuated the gas reservoir without further reaccretion (which is clearer in the thermal runs). Understanding the detailed fate of this gas remains an interesting avenue for future work.}

\textcolor{black}{The overall behaviour is consistent across L200m6, L200m7h and the L200m7 contours. However, in L200m7h, \textcolor{black}{the population of systems with very low $f_{\rm Edd}$ at $t_{\mathrm{q}}\lesssim200\,\mathrm{Myr}$ is less pronounced, with MQGs exhibiting a larger scatter across the full range of quenching times.}
Here, we average the Eddington ratio over the last $100\,\rm Myr$ of evolution, as the trends are clearer than for the instantaneous case, given the stochasticity of BH accretion.}

\textcolor{black}{These trends are consistent with the results of \citet{valentino_inprep}, who find that the most luminous sources (and therefore those with higher Eddington ratios) are preferentially associated with the youngest post-starburst systems. This is also in agreement with other galaxy formation models (such as {\sc Shark v2.0} -- \citealt{lagos24} -- shown in \citealt{valentino_inprep}), and suggests that BHs can continue accreting actively for several hundred Myr after quenching. In contrast, \citet{almaini25} find no clear evidence for enhanced AGN activity in the small fraction of MQGs detected in X-rays (less than $10\%$ in their sample), although their sample contains relatively few recently ($< 500\,\mathrm{Myr}$) quenched systems.}


\textcolor{black}{Overall}, we find a \textcolor{black}{clear} correlation between quenching and AGN feedback in {\sc COLIBRE}. This aligns with galaxy formation models such as \citet{lagos25}, which show that the AGN feedback prescription leaves clear imprints on MQG predictions. It also agrees with observations reporting AGN feedback signatures in MQGs \citep[e.g.][]{davies24, stevenson25, baker25b}. In {\sc COLIBRE}, AGN feedback drives outflows and shock-heats the intergalactic medium (IGM), suppressing gas cooling. This naturally explains the low molecular hydrogen reservoirs and reduced dust content arising from sputtering in hot gas \citep[see fig.~9 in][]{schaye25}, as discussed in \textcolor{black}{sec.~4.4 of \citetalias{chandro-gomez_inprep}}. \textcolor{black}{Appendix~\ref{appendix:images} further illustrates this process via images of individual MQGs, showing the removal of gas and dust by AGN-driven feedback bubbles.}

We now turn to \textcolor{black}{understanding} why some massive systems quench while others of similar mass remain star-forming.

\subsection{Environment (pre-selection or ``past'')}
\label{ssec:res2-environment} 

The fact that MQGs host BHs that release more energy relative to MSFGs raises the question of what is different between the two populations. Here we identify the origin of these differences. In this subsection, we present results for the thermal L200m6 simulation only, as the conclusions are unchanged for the hybrid L200m7h simulation (allowing for resolution effects).

\subsubsection{BH masses and star formation efficiencies}
\label{sssec:res2-environment-bhm} 

We begin by examining the BH–stellar mass relation in the left panel of Fig.~\ref{fig:mbh-mstars-eff-thermal}, focusing on the massive systems analysed in this section at $z=3$ (MQGs and MSFGs). \textcolor{black}{The background 2D maps} are colour-coded by \textcolor{black}{the median sSFR per bin}, showing that galaxies with more massive BHs tend to have lower sSFR values, consistent with AGN feedback driving quenching. The red and grey lines mark the median relations for MQGs and MSFGs, respectively, with shaded regions indicating the scatter. Passive systems clearly host more massive BHs, which in turn inject more feedback energy, as expected from the prescriptions in \S~\ref{sssec:colibre-agn-thermal} and \S~\ref{sssec:colibre-agn-hybrid}. In these models, the injected energy, whether in the form of a temperature increase (equation~\ref{eq:deltat-thermal}) or velocity kick (equation~\ref{eq:deltav-hybrid}), scales with BH mass (equations~\ref{eq:deltae-thermal}~and~\ref{eq:deltae-hybrid}). This trend is also evident in Appendix~\ref{appendix:corr} for MQGs and can be extended to MSFGs as well. Satellite MQGs (marked with black dots) host BHs with masses that lie predominantly above the MSF median, indicating that these systems are likewise quenched by BH physics (AGN feedback) rather than by environmental effects, \textcolor{black}{consistent with observations suggesting that, for massive galaxies at high-$z$, quenching is primarily linked to internal processes \citep[e.g.][]{lee15}.}

Unlike other models, such as {\sc IllustrisTNG} \citep{kurinchi-vendhan24}, where kinetic feedback is triggered once the Eddington ratio (a BH mass–dependent quantity) falls below a threshold, {\sc COLIBRE} does not impose a transition BH mass for AGN activity. Observationally, BH masses in high-$z$ quiescent systems remain difficult to constrain, although first estimates have emerged \citep[e.g.][]{carnall23b, ito25b}, making our predictions a useful benchmark for future surveys. \textcolor{black}{For comparison, we include the observational $M_{\mathrm{BH}}$--$M_{\star}$ \textcolor{black}{local} scaling relations from \citet{greene20} for early-type \textcolor{black}{and late-type galaxies, as well as for the full galaxy population, shown as dashed purple, dark blue and black lines}, respectively. We verify that all early MQGs in L200m6, not only those selected at $z=3$, broadly follow the \citet{greene20} relation for early-type systems. Therefore, consistent with some recent observational studies \citep{ito25b, li25}, MQGs in the model broadly follow the local relation and do not host overmassive BHs. \textcolor{black}{MSFGs instead tend to lie above the local relation for late-type systems. In the case of L200m7h, the BH masses of MQGs are lower (Fig.~\ref{fig:app-thermal-hybrid}) and skewed towards the local relation for late-type systems.}}

The right panel of Fig.~\ref{fig:mbh-mstars-eff-thermal} shows the star formation efficiency, defined as $\varepsilon_{\star}=M_{\star}/M_{\rm b}$ with $M_{\rm b}=(\Omega_{\rm b}/\Omega_{\rm m})\times M_{\rm halo}$, versus halo mass for only central massive systems. Scatter points are again colour-coded by sSFR, with a dashed line marking 10\% efficiency for reference. Galaxies with lower sSFR generally display higher efficiencies, indicating that quenched systems have been more effective at converting available gas into stars. By splitting the sample into MQGs and MSFGs, we see that MQGs are more efficient on average. Efficiency decreases with increasing halo mass. In {\sc COLIBRE}, MQG efficiencies remain below the $\gtrsim30\%$ values inferred from some spectroscopic studies \citep{carnall24, glazebrook24}, although these have been reported to decrease to $\sim10\%$ once environmental effects are considered \citep{jespersen25}. Our MQG efficiencies, however, are consistent with other theoretical predictions \citep{lagos25}. Higher efficiencies might be achieved if measured before the quenching epoch (considering all progenitors), since the efficiency can only decrease as stellar mass growth stalls.

Taken together, the results in Fig.~\ref{fig:mbh-mstars-eff-thermal} highlight systematic differences at the selection redshift: MQGs host more massive BHs and have higher \textcolor{black}{star formation} efficiencies than MSFGs. The key question is, therefore, what drives BHs to become more massive, or efficiencies to be higher, in MQGs. To answer this, we next trace the evolutionary histories of both populations using their main progenitors. 

\subsubsection{Evolutionary histories of MQGs}
\label{sssec:res2-environment-hist} 

Fig.~\ref{fig:agn-ev} shows the past evolution of several galaxy properties in lookback time, starting from the selection redshift $z=3$ backwards until $z=20$. \textcolor{black}{We restrict this analysis to central galaxies only (both MQGs and MSFGs) to minimise environmental effects and obtain a clearer comparison}. To avoid biases when comparing \textcolor{black}{central} MQGs and MSFGs and to ensure a truly stellar mass–matched sample, since MQGs are typically more massive, \textcolor{black}{we weight the central MSFG population such that its stellar mass distribution matches that of the central MQGs at $z=3$, following equation~(\ref{eq:w}) in \S~\ref{ssec:prop-mq-def}.} 


\textcolor{black}{The first row shows the evolution of stellar mass, $\mathrm{sSFR}=M_\star/\mathrm{SFR}$, and the gas metallicity. The latter is only available in the full simulation snapshots, and therefore has a lower temporal cadence.} The \textcolor{black}{second} row presents the evolution of BH mass, BH-injected energy, and the outflow mass rate of cool gas at $R_{\rm 200c}$. The \textcolor{black}{third} row shows the star formation efficiency, BHAR \textcolor{black}{($\dot{M}_{\rm accr}$ for the thermal AGN feedback model from equation~\ref{eq:maccr-rate})} averaged over 100 Myr (\textcolor{black}{$\min(100\,\mathrm{Myr},\Delta t_{\rm out})$}), and the inflow mass rate of cool gas at $0.3\,R_{\rm 200c}$. Median values are shown for MQGs (red) and MSFGs (grey), with shaded regions indicating the scatter. The vertical line marks the median quenching timescale $t_{\rm q}$, \textcolor{black}{defined as the lookback time from the selection redshift when the SFR last dropped below 10\% of its peak value (sec.~3.2 of \citetalias{chandro-gomez_inprep})}. \textcolor{black}{The gas metallicity is computed within the $50\,\rm pkpc$ fiducial aperture as the gas-mass-weighted linear sum of the diffuse (\textcolor{black}{excluding metals locked into dust}) oxygen-to-hydrogen ratio of gas}. The mass flow rates are computed as described in \S~\ref{sec:prop}, focusing on cool gas with temperatures $10^{3}\,\mathrm{K}<T_{\rm gas}<10^{5}\,\mathrm{K}$, since this is the dominant component. It traces both the cold streams that supply inflows to galaxies, and the possible neutral/ionised outflows reported observationally \citep[e.g.][]{d'eugenio24}. The main conclusions from these panels are:

\begin{figure*}
\centering
\includegraphics[trim={0 0 0 0},clip, width=\textwidth]{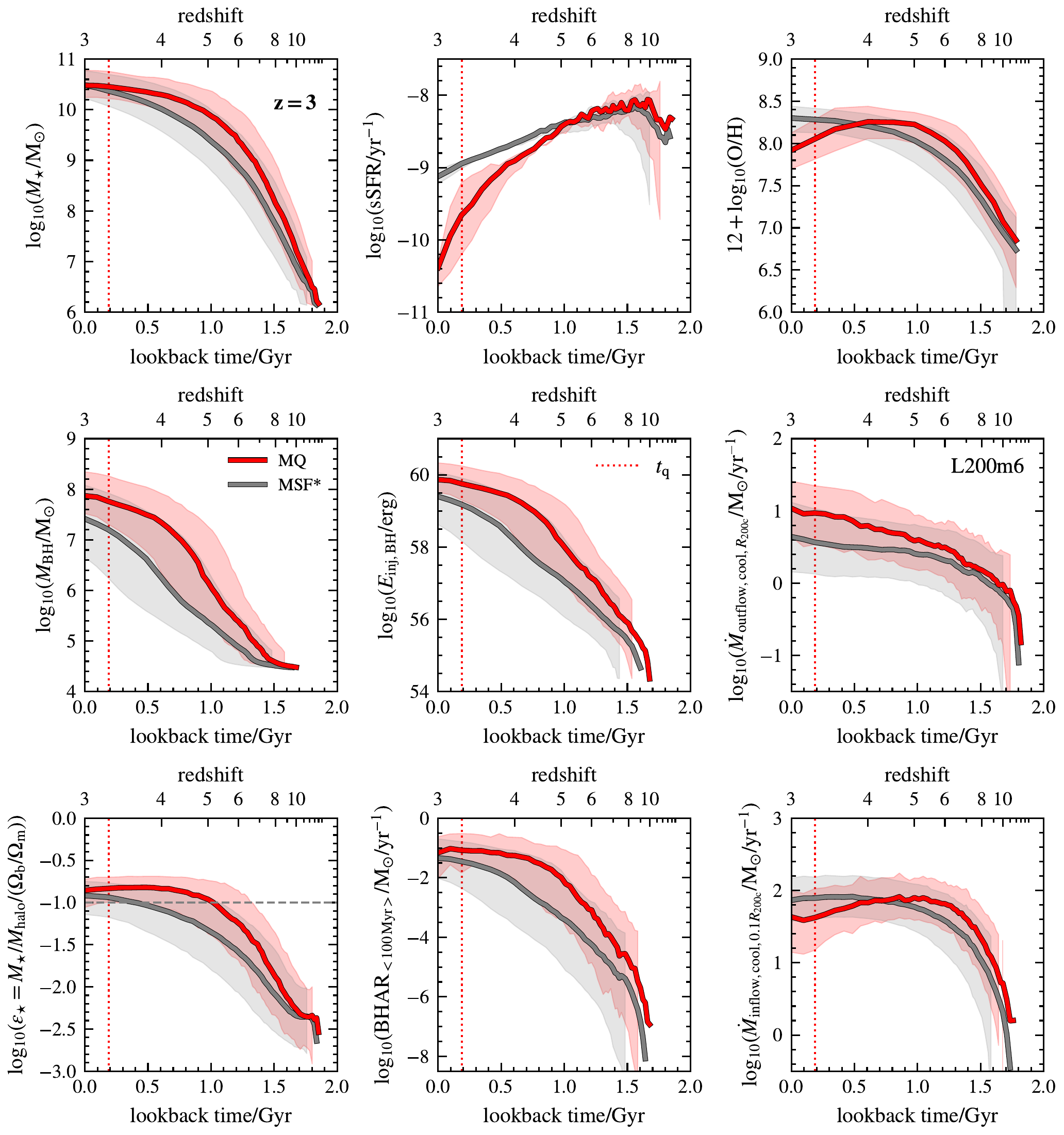}
\caption{Evolution of \textcolor{black}{galaxy}- and \textcolor{black}{BH}-related properties as a function of lookback time (with redshift indicated by the top $x$-axis) for \textcolor{black}{all the central $z=3$ MQGs and (stellar-mass weighted) MSFGs} galaxies. \textcolor{black}{\textit{First row}: stellar mass (\textit{left}) $\mathrm{sSFR}=M_\star/\mathrm{SFR}$ (\textit{middle}), and gas metallicity (gas-mass-weighted linear sum of the diffuse oxygen-to-hydrogen ratio of gas) only for the full snapshots (\textit{right}).} \textit{\textcolor{black}{Second} row}: BH mass (\textit{left}) cumulative BH-injected energy (\textit{middle}), and outflow rate of cool gas ($10^{3}\,\mathrm{K}<T_{\rm gas}<10^{5}\,\mathrm{K}$) at $R_{\rm 200c}$ (\textit{right}). \textit{\textcolor{black}{Third} row}: star formation efficiency, $\varepsilon_{\star}=M_{\star}/M_{\rm halo}/(\Omega_{\rm b}/\Omega_{\rm m})$ (\textit{left}), BHAR \textcolor{black}{($\dot{M}_{\rm accr}$ for the thermal AGN feedback model)} averaged over 100 Myr (\textit{middle}), and inflow rate of cool gas ($10^{3}\,\mathrm{K}<T_{\rm gas}<10^{5}\,\mathrm{K}$) at $0.1\,R_{\rm 200c}$ (\textit{right}). \textcolor{black}{Note that the averaged quantity is computed over $\min(100\,\mathrm{Myr}, \Delta t_{\rm out})$, where $\Delta t_{\rm out}$ is the time interval between consecutive simulation outputs.} Solid lines show the median values for MQGs (red) and MSFGs (grey) selected as in \S~\ref{ssec:prop-mq-def} at $z=3$, with shaded regions indicating the 16th and 84th percentile range. The vertical red dotted line marks the median quenching timescale, $t_{\rm q}$, of the MQGs. In the bottom left panel, the dashed horizontal line marks a reference value of 10\% star formation efficiency.} 
\label{fig:agn-ev} 
\end{figure*}

\begin{figure*}
\centering
\includegraphics[trim={0 0 0 0},clip, width=\textwidth]{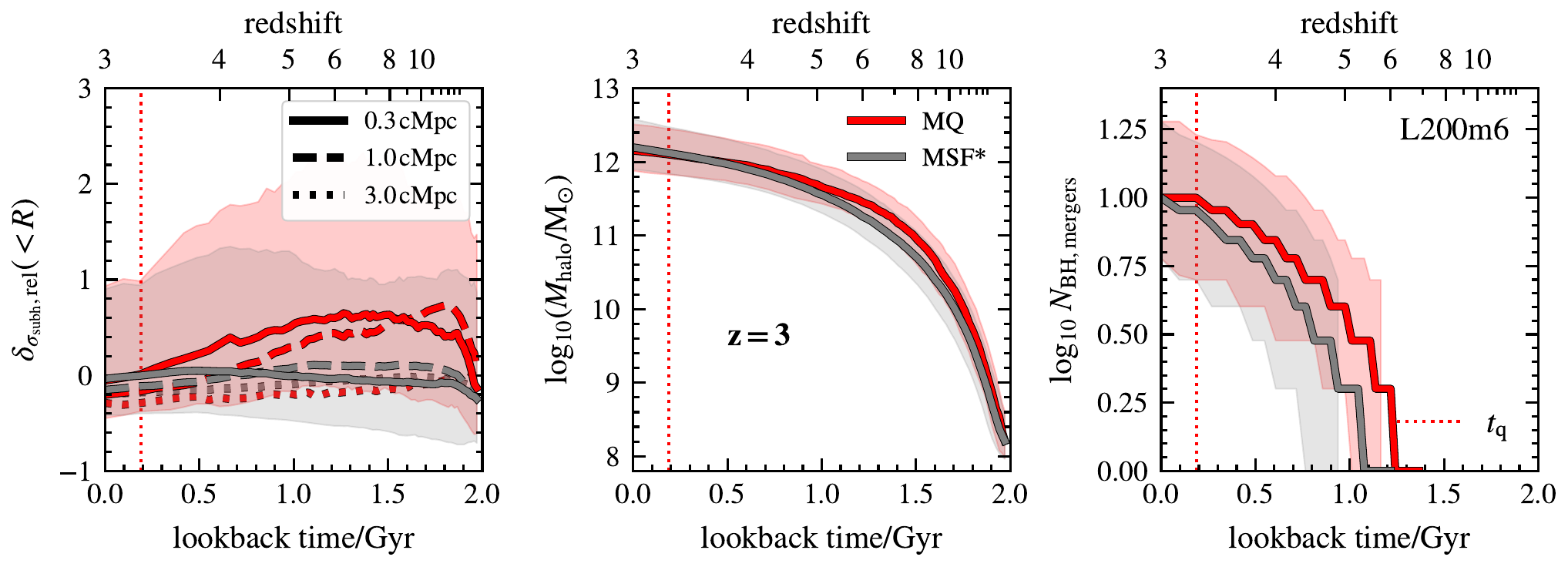}
\caption{Evolution of environment-related galaxy properties as a function of lookback time (with redshift indicated by the top $x$-axis) for \textcolor{black}{all the central $z=3$ MQGs and (stellar-mass weighted) MSFGs}. $\delta_{\rm \sigma_{subh},rel}(<R)$ deviation from the mean subhalo density contrast, as defined in \S~\ref{ssec:prop-delta-sigma}, in a sphere of radius $R=0.3, 1\ \mathrm{or}\ 3.0\, \mathrm{cMpc}$ (\textit{left panel}); halo mass only for central galaxies (\textit{middle panel}); and number of BH mergers (\textit{right panel}). Solid lines show the median values for MQGs (red) and MSFGs (grey) selected as in \S~\ref{ssec:prop-mq-def} at $z=3$, with shaded regions indicating the 16th and 84th percentile range (for the left panel, this shading corresponds only to the $R = 0.3\,\mathrm{cMpc}$ curves. The vertical red dotted line marks the median quenching timescale, $t_{\rm q}$, of the MQGs.} 
\label{fig:agn-ev-env} 
\end{figure*}

\begin{itemize}
    \item \textcolor{black}{\textbf{Stellar mass}: MQGs assemble their stellar mass earlier and more rapidly than MSFGs, after which their growth slows down due to AGN feedback, which suppresses further star formation until their stellar masses converge at the selection redshift.}
    \item \textcolor{black}{\textbf{sSFR}: MQGs exhibit higher sSFRs (starburst phases) at early times ($z\gtrsim5$), consistent with their faster stellar mass assembly. At later times, their sSFR drops sharply as the galaxies become quiescent due to AGN feedback.}
    \textcolor{black}{\item \textbf{Gas metallicity (O/H)}: MQGs are initially more metal-rich, likely due to enhanced star formation activity. However, at $z\lesssim4$, their gas metallicity decreases and eventually becomes lower than that of MSFGs by the selection time. This decline is likely driven by reduced star formation (limiting metal production) and feedback-driven outflows that expel enriched material \citep[e.g.][]{baron18}.}
    \item \textbf{BH mass}: MQG BHs are seeded slightly earlier, reflecting their host FOF/haloes being slightly more massive (\S~\ref{ssec:colibre-agn}). They also grow at a much faster rate for $z\lesssim8$. The earlier seeding and faster growth of BHs in quenched systems is consistent with the behaviour reported in \citet{kurinchi-vendhan24}. \textcolor{black}{In MQGs, stellar mass is assembled earlier than BH mass, implying that their BHs remain slightly undermassive \textcolor{black}{and, in L200m6, lie closer to the local BH--stellar mass scaling relation for all and late-type galaxies from \citet{greene20} before selection.}}
    \item \textbf{BH-injected energy}: following the same trend, MQGs inject more energy on average (\S~\ref{ssec:res2-agn-feedback}). This is consistent with the expectation that AGN energy drives their quenching, which is captured in the {\sc COLIBRE} feedback prescription (\S~\ref{ssec:colibre-agn}). 
    \item \textbf{Outflowing mass rate}: we trace the gas outflowing at $R_{\rm 200c}$, which we use to define the halo boundary. This allows us to track material expelled from the halo. As MQG BHs grow more massive and inject more energy, they drive stronger outflows, particularly by $z\lesssim6$, \textcolor{black}{coinciding with the divergence in sSFR between MQGs and MSFGs}. These powerful outflows, driven by the AGN, suppress the local cooling rate of the circumgalactic medium (CGM) and contribute to quenching. They are the primary mechanism for heating and removing the galaxy's gas reservoir \textcolor{black}{(as seen in the images in Figs.~\ref{fig:images0}--\ref{fig:images2} for individual systems)}, which directly leads to the low molecular hydrogen and dust fractions discussed in sec.~4.4 of \citetalias{chandro-gomez_inprep}. \textcolor{black}{This picture is also supported observationally by a recent detection of a powerful AGN-driven outflow in a quasar that has been proposed as a candidate progenitor of a MQG \citep{liu26}.}
    \item \textbf{Star formation efficiency ($\varepsilon_{\star}$)}: MQGs reach higher efficiencies, remaining above the 10\% reference value for longer (for main progenitors at least). Efficiencies are higher before the quenching event and decline afterwards as stellar mass growth saturates.
    \item \textbf{BHAR}: we employ a $\rm 100\,Myr$ time-averaged BHAR, \textcolor{black}{computed over $\min(100\,\mathrm{Myr},\Delta t_{\rm out})$ for cases where the interval between consecutive simulation outputs is shorter than $100\,\rm Myr$. This} smoothes over the short timescale fluctuations in the instantaneous value. The shapes of the curves mirror those of the BH-injected energy evolution, with MQGs offset earlier due to earlier seeding. Importantly, BHAR typically peaks before quenching, after which accretion reaches a plateau or drops. This is as we would expect: rapid accretion triggers stronger feedback, which quenches the system by driving outflows and heating the surrounding gas, suppressing further efficient accretion. A similar feature is reported in {\sc FLARES}, a set of zoom-ins using the {\sc EAGLE} model \citep{lovell23}.
    \item \textbf{Inflowing mass rate}: we trace the gas inflowing at $0.1R_{\rm 200c}$, which is expected to follow cold streams feeding the galaxy, though it may also include some recycled gas from feedback \citep{mitchell20, wrightr24}. MQGs initially show larger inflows of cool material, but once their BHs become sufficiently powerful, these inflows decline, falling below those of MSFGs. This prevents replenishment of the gas reservoir and reinforces quiescence as a direct consequence of AGN heating and gas expulsion. As these are mass-matched samples, we might expect their median values to cross over at some point in the past.
\end{itemize}

Our results make clear that the evolutionary histories of MQGs and MSFGs diverge between $z=3$ and earlier times: quiescent systems host more massive, energetic BHs and achieve higher stellar-to-halo mass ratios due to greater early accretion of cool gas. This raises the question of what drives the differences in the first place.

\subsubsection{Environment shaping MQGs}
\label{sssec:res2-environment-env} 

We examine the environment and merger history of MQGs in Fig.~\ref{fig:agn-ev-env}. The environment is characterised by $\delta_{\rm \sigma_{subh},rel}(<R)$, the deviation from the mean subhalo density contrast in different spherical apertures ($0.3,\ 1.0,\ \mathrm{or}\ 3.0\,\mathrm{cMpc}$) introduced in \S~\ref{ssec:prop-delta-sigma}, and by the halo mass. The merger history is traced by the number of BH mergers, which serve as a `proxy' for mergers, since, by construction, these only occur when both merging galaxies are massive enough to host seeded BHs (\S~\ref{ssec:colibre-agn}). These three tracers (subhalo density contrast, halo mass, and mergers) correlate as shown in Appendix~\ref{appendix:corr} for MQGs, with this correlation extending to MSFGs as well.

Fig.~\ref{fig:agn-ev-env} shows that \textcolor{black}{central MQGs and MSFGs typically occupy very similar environments at the selection time, with comparable host halo masses and merger activity. However, the differences become really pronounced at earlier epochs prior to quenching ($z\gtrsim4$ in this case), where MQGs are found in systematically more overdense regions. This is particularly evident} on local ($0.3\,\mathrm{cMpc}$, halo) and intermediate ($1.0\,\mathrm{cMpc}$, local cosmic web) scales, \textcolor{black}{while it largely disappears on larger scales ($3.0\,\mathrm{cMpc}$)}. The environmental differences are significant beyond $z \gtrsim 4$ but smooth out by the selection time, being more pronounced at larger scales, the earlier we consider. This likely reflects our use of a subhalo density contrast, highlighting that these neighbouring subhalos are eventually accreted. Plus, rather than being the primary quenching mechanism, the higher merger rate in MQGs is another consequence of their dense environment, serving to accelerate mass assembly, triggering star formation bursts and potentially fueling the rapid BH growth that ultimately powers quenching feedback. This is consistent with observational evidence that merger fractions increase with local density at such high redshifts \citep{shibuya25}. Although mergers are more common in MQGs, they do not produce morphological differences between MQGs and MSFGs (\textcolor{black}{sec.~4.6 of \citetalias{chandro-gomez_inprep}}).


Overall, MQGs are more likely to reside in overdense regions, particularly at earlier times in their evolutionary history. Their early position within the cosmic web and host halo masses imply deeper potential wells and stronger gas inflows, fuelling both star formation and BH accretion. This, in turn, enables higher stellar masses, faster supermassive BH growth, and ultimately more powerful feedback, producing outflows that quench them more efficiently. These are the trends evident in Fig.~\ref{fig:agn-ev-env}. 

The correlation between environment and quenching has been widely reported in the literature. For instance, \citet{kurinchi-vendhan24} find in {\sc IllustrisTNG} that large-scale structure plays a key role in setting galaxy evolutionary paths, a result also seen in {\sc Astrid} \citep{weller25}. \textcolor{black}{{\sc GAEA} finds MQGs across a broad range of halo masses, although quenched systems preferentially occupy the most massive haloes and are shifted towards higher halo masses compared to MSFGs \citep{delucia26}.} In {\sc Thesan}, which includes radiative transfer and enhanced BH growth, \citet{chittenden25} show that MQGs exhibit faster BH accretion and stronger environmental imprints, especially on small to intermediate scales, consistent with our {\sc COLIBRE} results. Similarly, \citet{szpila25} report that MQGs in {\sc Simba-C} occupy more overdense regions at earlier times, before the selection, which enhances the efficiency of feedback coupled with early and rapid BH growth. However, by the selection time, they are found in less dense environments than their massive star-forming counterparts, because in their model, powerful kinetic jets that quench the galaxy require low BHARs, which only occur when the surrounding environment is not too dense. We note that \citet{kimmig25} find the opposite trend: MQGs occupy less overdense regions, particularly at local scales, even before selection. These systems form in local, rather than global, nodes associated with weaker filaments, which AGN feedback can easily disrupt.

Observational constraints remain sparse, but emerging results suggest a consistent picture of MQGs residing in overdense environments \citep{jin24, ito25, degraaff25, kawinwanichakij25, baker25c, mcconachie25, kakimoto26}, with the caveat that this observed trend will be influenced by survey biases, which target overdense regions. \textcolor{black}{In contrast, \citet{binh25} report that MQGs reside in environments that are, on average, comparable in density to those of MSFGs.} 



\subsection{Rejuvenation (post-selection or ``future'')}
\label{ssec:res2-rejuvenation} 

The final stage of our journey to understand the physical drivers that shape MQGs focuses on examining their future evolution, a prediction that is possible in simulations but not in observations. We analyse some properties of the main descendants from {\sc HBT-HERONS} of all the MQGs and MSFGs (without applying any stellar mass binning) in Fig.~\ref{fig:rej}, where we look at the number fraction of galaxies as a function of the age of the Universe. We compare results from the thermal L200m6 simulation (top panel) with those from the hybrid L200m7h simulation (bottom panel).

\subsubsection{Number fraction of surviving main progenitors}
\label{ssssec:res2-rejuvenation-surv} 

\begin{figure}
\centering
\includegraphics[trim={0 0 0 0},clip, width=0.48\textwidth]{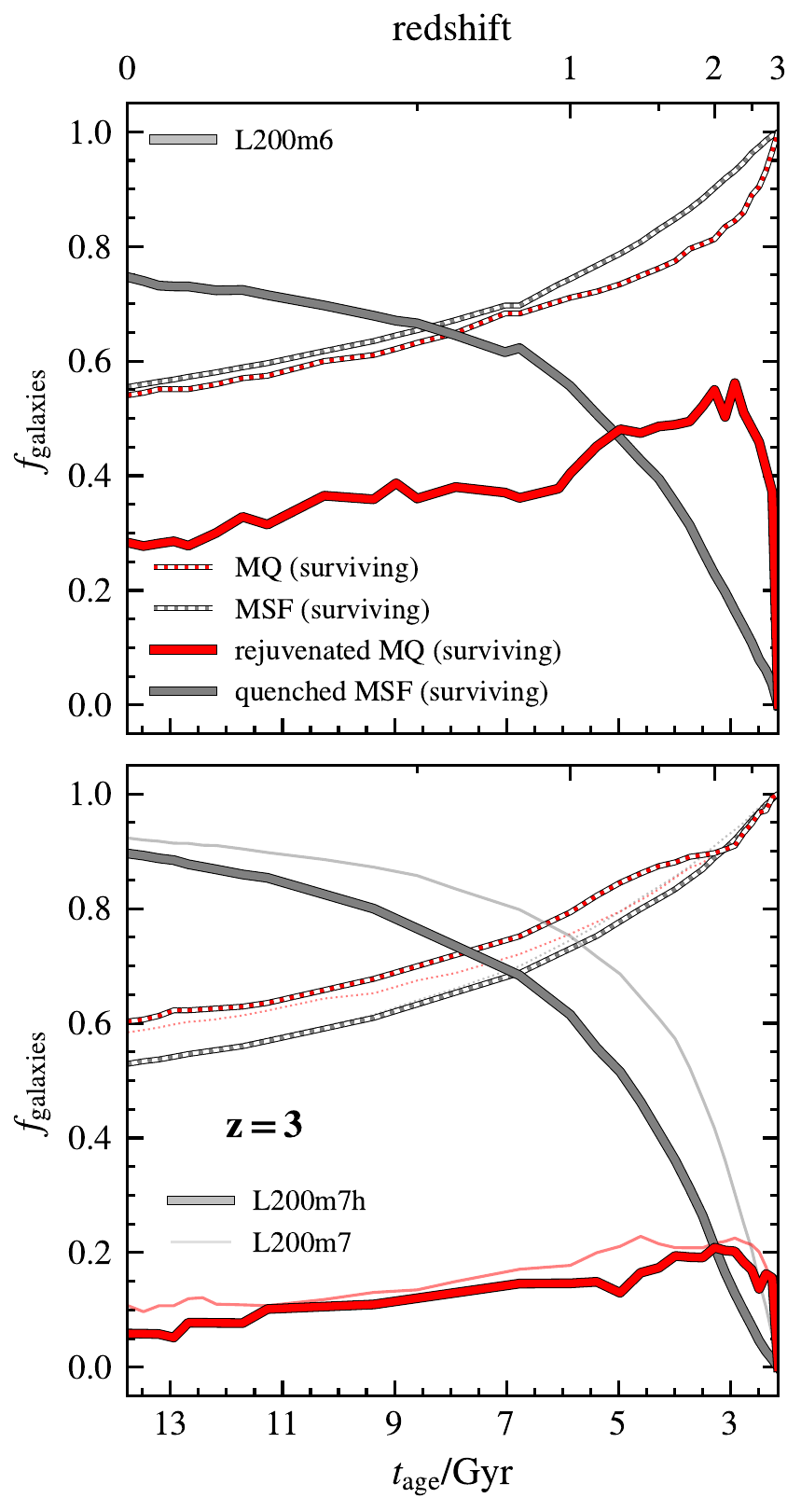}
\caption{Evolution of the number fraction of galaxies as a function of the age of the Universe from $z=3$ to $z=0$ (redshift by the top $x$-axis). Solid red lines show galaxies that are MQGs at $z=3$ and subsequently rejuvenate (they are star-forming at the corresponding time on the $x$-axis), while solid grey lines show galaxies that are MSFGs at $z=3$ and later quench (they are quenched at the corresponding time). Dotted red and grey lines indicate galaxies whose main progenitors are the MQGs and MSFGs selected at $z=3$, respectively, at each given time. Results are shown for the thermal L200m6 simulation (\textit{top panel}) and the hybrid L200m7h simulation (\textit{bottom panel}). Narrow solid lines in the background show the L200m7 results, included to illustrate the AGN feedback model effect.}
\label{fig:rej} 
\end{figure}

First, we consider the number fraction of passive systems selected at $z=3$ that survive as main descendants; that is, galaxies that have not merged into another galaxy. \textcolor{black}{Note that the exact values depend on the definition adopted for ``massive quenched'' galaxies, although the overall trends remain similar.} This fraction is shown in Fig.~\ref{fig:rej} at each given time as a red dotted line. By $z=0$, around \textcolor{black}{54}\% of these systems are still detected in the simulation (either as centrals or satellites). The numbers are similar for MSFGs, shown with grey dotted lines. 
This fraction of massive galaxies without experiencing mergers is consistent with the inside-out growth scenario, where systems grow primarily through minor mergers until $z=0$, eventually becoming the elliptical galaxies observed locally \citep{bezanson09}. 67\% (71\%) \textcolor{black}{of the galaxies whose main progenitors are the $z=3$ MQGs are central galaxies at $z=0$} in the thermal (hybrid) simulation. Future work will investigate whether these correspond to the brightest cluster galaxies. The thermal and hybrid simulations produce broadly similar results in terms of how these systems survive down to $z=0$.

Comparisons with other simulations show some variation, considering the different ``massive quenched'' selection criteria. In {\sc IllustrisTNG}, \citet{hartley23} report that only 1 galaxy out of five selected MQGs at $z\approx4.2$ in {\sc TNG300} has a main descendant that survives to $z=0$ (20\%). \citet{weller25} find $\approx60\%$ at $z=0$, with $30\%$ identified as brightest cluster galaxies when selecting the sample at $z=3$ in {\sc TNG100} and {\sc TNG300}. These values are broadly consistent with the survival fraction we find in {\sc COLIBRE}, given the comparable trends in terms of environment (\S~\ref{ssec:res2-environment}). In the {\sc Magneticum} simulations, \citet{remus25} report that by $z=2$, $\approx20\%$ of MQGs selected around $z=3$ are accreted onto more massive structures (either as satellite galaxies or by having merged).

\subsubsection{Number fraction of rejuvenated galaxies}
\label{sssec:res2-rejuvenation-rej} 

We also investigate in Fig.~\ref{fig:rej} rejuvenation events.
We compute the number fraction of quenched systems at $z=3$ that temporarily exceed the sSFR threshold defined in \S~\ref{ssec:prop-mq-def} at a given redshift $z<3$, relative to those that survive at each redshift (not having merged). Rejuvenation peaks around \textcolor{black}{56}\% at \textcolor{black}{$z=2.25$} for the thermal L200m6 simulation, compared to only around \textcolor{black}{21}\% for the hybrid L200m7h \textcolor{black}{at $z=2$}. Initially, it takes $\approx 1\,\rm Gyr$ for the ISM to sustain star formation again in some systems, leading to an increase in the rejuvenation fraction from $z=3$ to $z\approx2$. Later, in these systems, presumably cooling and renewed gas inflow allow temporary star formation. However, BHs eventually re-establish strong feedback, reducing the rejuvenation fraction again from $z\approx2$ to $z=0$. Although we do not quantify this hypothesis nor the role of mergers in supplying gas for star formation, rejuvenated galaxies generally host lower-mass BHs, indicating weaker quenching that allows star formation to resume temporarily. In massive systems, strong BH feedback ultimately ensures most descendants remain quiescent as can be seen from the increasing fraction of MSFGs selected at $z=3$ that quench with time by $z<3$ (solid grey lines). The hybrid simulation results broadly follow those of L200m7 (background lines), pointing to the AGN feedback model effect analysed previously (\textcolor{black}{sec.~4.1 of \citetalias{chandro-gomez_inprep}}): AGN feedback is stronger and burstier in the m7 resolution runs, preventing star formation from resuming. This explains why more MSFGs quench by $z=0$ as well (solid grey line).

These trends are broadly consistent with other galaxy formation models \citep[considering the different MQ selection;][]{szpila25, weller25, remus25}, although they typically report slightly lower rejuvenation fractions ($\approx30\%$ at $z\gtrsim2$, dropping to $<5\%$ at lower redshift); this is likely due to stronger AGN feedback. Observationally, as explained before, detecting rejuvenation features in possible MQGs descendants is difficult because the SED-fitting process struggles to recover such features in SFHs, although some evidence exists from local counterparts \citep{yi05}.
Cumulatively, around 83\% of the $z=3$ MQGs in L200m6 experience rejuvenation at some point by $z=0$ , although this fraction depends on the snapshot cadence used to identify such events. Among the systems that survive as main descendants in this simulation, the fraction rises to around 98\%, with rejuvenation episodes slightly above the sSFR threshold. Thus, rejuvenation appears to be a very common phase in the evolution of these systems.

\section{Conclusions}
\label{sec:conclusions}

\textcolor{black}{Recent \textit{JWST} surveys have revealed that Massive Quenched Galaxies (MQGs; defined as galaxies with $M_{\star}>10^{10}\,\mathrm{M_{\odot}}$ and $\mathrm{sSFR}<0.2/t_{\rm age}$) \textcolor{black}{at $z\geq2$} are more abundant than previously inferred. Building on the results of \citetalias{chandro-gomez_inprep}, which showed that the new {\sc COLIBRE} hydrodynamical simulations reproduce the observed MQG population broadly well across a range of properties, particularly number densities and stellar mass functions, we investigated how MQGs assemble their mass and quench so rapidly within the {\sc COLIBRE} model. In particular, we focused on the role of AGN feedback, comparing the thermal implementation in the L200m6 simulation with the hybrid implementation in L200m7h, as well as on the role of environment. To do this,} we traced the evolutionary histories \textcolor{black}{of MQGs}, focusing \textcolor{black}{mainly} on galaxies selected at $z=3$. Our key conclusions are:

\begin{itemize}
    \item \textbf{AGN feedback drives quenching}: Both thermal and hybrid (thermal+jet) AGN feedback models quench MQGs (\S~\ref{ssec:res2-agn-feedback}). \textcolor{black}{Small-volume simulation runs clearly show that, without AGN feedback, MQGs cannot form at $z>1$ (\S~\ref{sssec:res2-agn-feedback-ndens}). However,} the hybrid AGN feedback model produces less quenching due to \textcolor{black}{(i) a slower BH growth at early times and (ii) because the modelled jets, which dominate the energy budget, take longer to impact the surrounding gas and trigger quenching ($z\lesssim3$)}. This clearly highlights that the timing and power of early AGN feedback are critical for rapid quenching. \textcolor{black}{In addition, \citet{chaikin26} demonstrate the importance of super-Eddington accretion at high redshift for the emergence of these systems. Furthermore, the central/satellite hierarchy appears to play only a minor role, with internal BH-driven processes dominating over environmental mechanisms, since satellite MQGs also host massive BHs (Fig.~\ref{fig:mbh-mstars-eff-thermal}).}
    \item \textcolor{black}{\textbf{AGN properties}}: {\sc COLIBRE} also predicts that some MQGs host quasars (\textcolor{black}{luminous} AGNs with $L_{\rm bol}\gtrsim10^{45}\,\mathrm{erg\,s^{-1}}$), broadly consistent with recent observations, although such systems remain rare at extreme redshifts ($z\gtrsim5$). More luminous AGNs are preferentially associated with more recently quenched galaxies (\S~\ref{sssec:res2-agn-feedback-obs}).
    \item \textbf{Environment matters}: \textcolor{black}{MQGs and MSFGs occupy similar environments \textcolor{black}{on scales much larger than the galaxies themselves} at the selection time, with comparable halo masses and merger activity. However, MQGs are found in significantly more} overdense regions before the selection time \textcolor{black}{and the quenching event} \textcolor{black}{on scales of $0.3 - 1.0\,\mathrm{cMpc}$} (\S~\ref{ssec:res2-environment}); enabling earlier gas accretion, BH growth and more powerful AGN feedback. This explains their accelerated evolution and enhanced star formation efficiencies.
    \item \textbf{Rejuvenation}: Rejuvenation episodes are common, peaking at $z\approx2$ with up to \textcolor{black}{56}\% of MQGs experiencing renewed star formation in the thermal AGN feedback simulation and ~20\% in the hybrid AGN feedback simulation (\S~\ref{ssec:res2-rejuvenation}). About \textcolor{black}{54\% (60\%) of MQGs, respectively, survive as the main progenitors of $z=0$ galaxies.}
\end{itemize} 

\textcolor{black}{Together with \citetalias{chandro-gomez_inprep}, this series aims to provide robust predictions for the MQG population within the {\sc COLIBRE} framework, helping to ease the tension between observations and simulations (\citetalias{chandro-gomez_inprep}) while offering insight into the physical mechanisms that shape their origin and evolution (this Paper~II). Although important caveats remain on both the observational side (e.g. systematics and sample selection) and the simulation side (e.g. uncertainties in BH growth and AGN feedback modelling), the main conclusions of this series remain robust within these uncertainties and provide a consistent framework for interpreting MQG formation and evolution.} 

Looking ahead, {\sc COLIBRE} opens several promising avenues for follow-up studies. One direction is to further investigate the connection between the halo assembly and MQG evolution. Predictions for galaxy clustering will also be valuable, as they can help establish links between MQGs and potential progenitor populations, such as highly star-forming galaxies. Upcoming observations leveraging the synergy of \textit{JWST} and ALMA will allow detailed characterisation of the stellar populations and clustering of both populations, providing a way to test these connections \citep{mcalpine19, hygate23, long23, bodansky25}. Moreover, MQGs may also be linked to other high-$z$ populations probed by \textit{JWST}, including early \textcolor{black}{luminous galaxies at $z\gtrsim10$ \citep{bunker23}} or the so-called “Little Red Dots” \citep[LRDs;][]{kokorev24, matthee24}, as well as to descendant populations such as the most massive elliptical galaxies in the local Universe or the observed rare class of compact, massive early-type “relic” galaxies, \textcolor{black}{identified observationally \citep{spiniello21} and recently investigated in {\sc COLIBRE} by \citet{zemsch26}}. Additional work \textcolor{black}{will} test \textcolor{black}{{\sc COLIBRE} predictions for chemical abundances of MQGs (O'Connor et al., in prep.)}, \textcolor{black}{motivated by observations of $\alpha$-enhancement \citep{hamadouche26} and their interpretation through chemical evolution modelling \citep{oconnor26}}.


\section*{Acknowledgements}

\textcolor{black}{We thank the anonymous referee for their constructive feedback.} ACG acknowledges Research Training Program and ICRAR scholarships. ACG acknowledges support for this project by the University of Western Australia via a Research Collaboration Award. This work used the DiRAC@Durham facility managed by the Institute for Computational Cosmology on behalf of the STFC DiRAC HPC Facility (\url{www.dirac.ac.uk}). The equipment was funded by BEIS capital funding via STFC capital grants ST/K00042X/1, ST/P002293/1, ST/R002371/1 and ST/S002502/1, Durham University and STFC operations grant ST/R000832/1. DiRAC is part of the National e-Infrastructure. This project has received funding from the Netherlands Organization for Scientific Research (NWO) through research programme Athena 184.034.002. WMB would like to acknowledge support from DARK via the DARK Fellowship. This work was supported by a research grant (VIL54489) from VILLUM FONDEN. ABL acknowledges support by the Italian Ministry for Universities (MUR) program “Dipartimenti di Eccellenza 2023-2027” within the Centro Bicocca di Cosmologia Quantitativa (BiCoQ), and support by UNIMIB’s Fondo Di Ateneo Quota Competitiva (project 2024-ATEQC-0050). \textcolor{black}{EC acknowledges support from STFC consolidated grant ST/X001075/1.} SP acknowledges support by the Austrian Science Fund (FWF) through grant-DOI: 10.55776/V982. JT acknowledges support of a STFC Early Stage Research and Development grant (ST/X004651/1). Minor typos, grammar and spelling mistakes were identified with the assistance of ChatGPT-4o\footnote{\url{openai.com}} when preparing this document. No passages of text or structural outlines for this paper were created with the help of any large language models.

\section*{Data Availability}

The data underlying this article will be provided upon reasonable request to the corresponding author.



\bibliographystyle{mnras}
\bibliography{colibre-mqg} 




\appendix

\section{Correlations for MQGs}
\label{appendix:corr}

\begin{figure*}
\centering
\begin{subfigure}[b]{0.48\textwidth}
   \includegraphics[trim={0 0 0 0},clip, width=\textwidth]{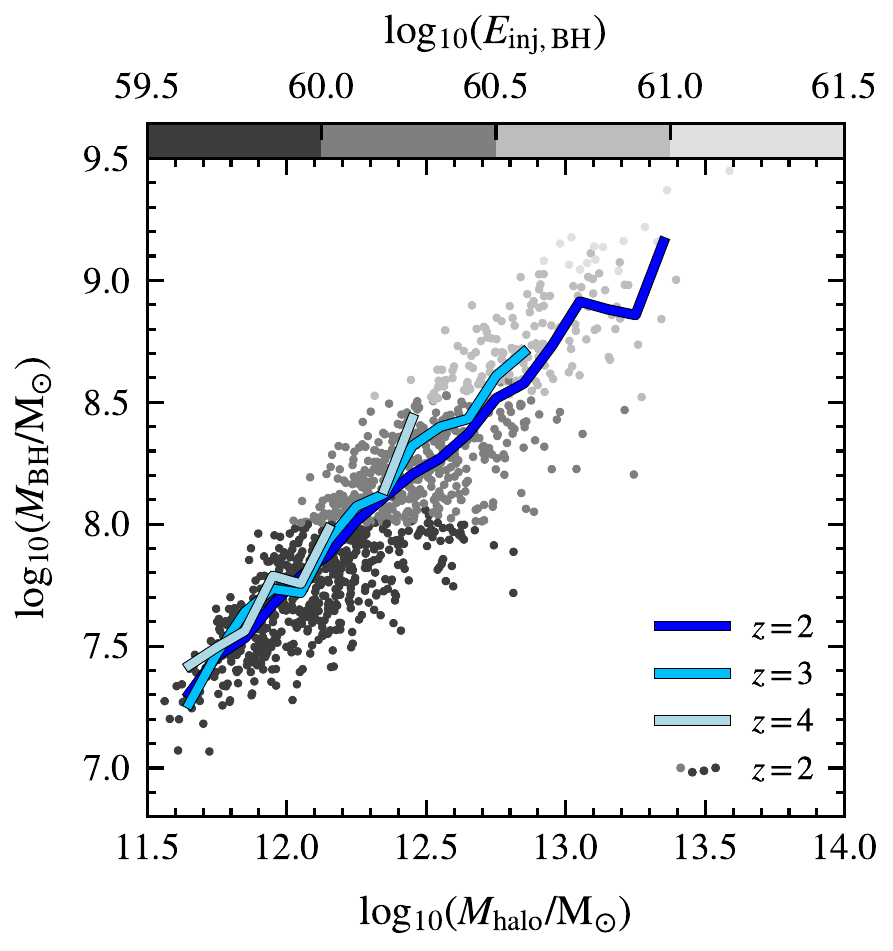}
\end{subfigure}
\begin{subfigure}[b]{0.48\textwidth}
   \includegraphics[trim={0 0 0 0},clip, width=\textwidth]{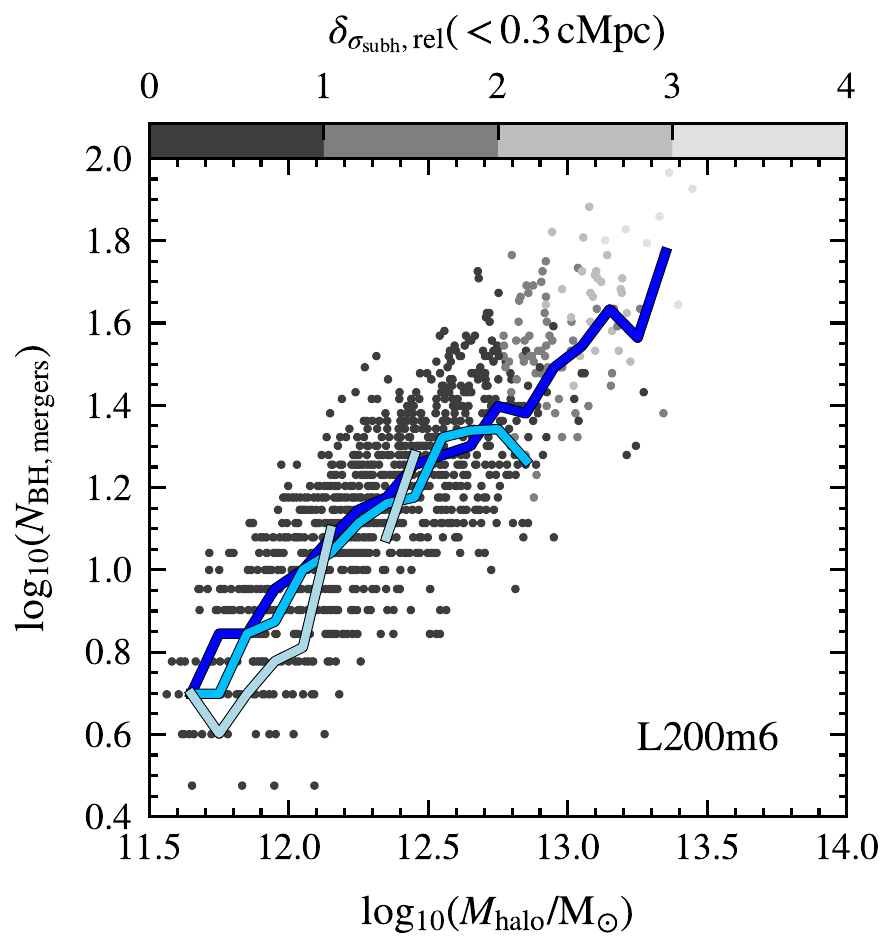}
\end{subfigure}
\caption{BH mass versus halo mass, colour-coded by BH-injected energy (\textit{\textcolor{black}{left panel}}); and number of BH mergers versus halo mass, colour-coded by $\delta_{\rm \sigma_{subh},rel}(<R)$ deviation from the mean subhalo density contrast, as defined in \S~\ref{ssec:prop-delta-sigma}, within a sphere of $R=0.3\,\mathrm{cMpc}$ (\textit{right panel}) for \textcolor{black}{central} MQGs \textcolor{black}{in L200m6}. Solid lines show the median values at different redshifts ($z=2,3,4$), while scatter points represent individual MQGs selected at $z=2$, colour-coded by the properties indicated above.} 
\label{fig:corr-env} 
\end{figure*}

\textcolor{black}{We present two environment-related correlations for central MQGs predicted by the {\sc COLIBRE} model. The left panel of Fig.~\ref{fig:corr-env} shows the relation between BH mass and halo mass, with median trends indicated by different colours for galaxies selected at different redshifts ($2 \leq z \leq 4$) in the L200m6 simulation. Background scatter points correspond to individual galaxies selected at $z=2$, colour-coded by the cumulative energy injected by the BH through feedback. Denser environments are clearly associated with more massive BHs and larger feedback energies, consistent with the trends discussed in \S~\ref{ssec:res2-environment}. The right panel of Fig.~\ref{fig:corr-env}, following the same format, compares the three environmental tracers used throughout this work: halo mass, number of BH mergers, and $\delta_{\rm \sigma_{subh},rel}(<R)$, defined as the deviation from the mean subhalo density contrast (\S~\ref{ssec:prop-delta-sigma}). All three tracers show strong correlations with one another.}

\section{Evolutionary histories and galaxy images for individual MQGs}
\label{appendix:images}

\textcolor{black}{This appendix further illustrates that AGN feedback is responsible for quenching, as discussed in \S~\ref{ssec:res2-agn-feedback}, presenting individual examples of MQGs at different stages after quenching.}

\begin{figure*}
\centering
\includegraphics[width=0.98\textwidth]{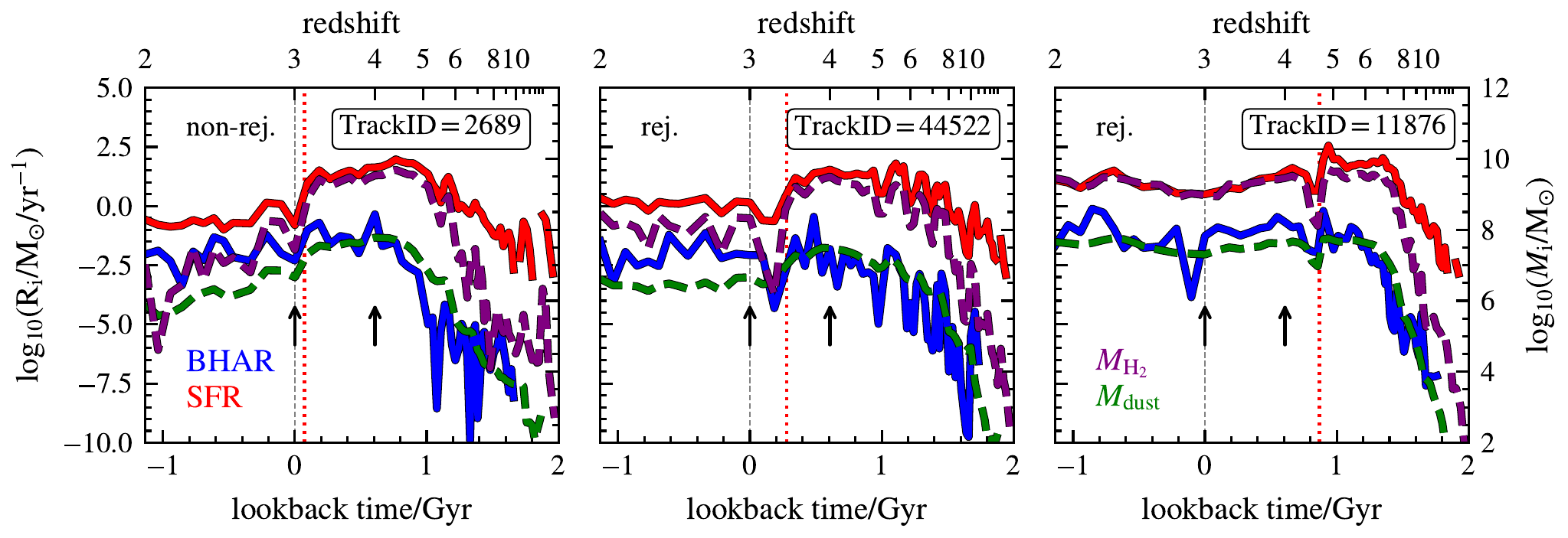}
\caption{\textcolor{black}{Evolution of galaxy- and BH-related properties as a function of lookback time (with redshift indicated by the top $x$-axis) for three individual central $z=3$ MQGs in the L200m6 simulation, shown in separate panels. We indicate the corresponding TrackID from the {\sc SOAP} catalogue in the top right corner of each panel. Solid lines show the SFR (red) and instantaneous BHAR, \textcolor{black}{$\dot{M}_{\rm accr}$ for the thermal AGN feedback model,} (blue); using the scale on the left $y$-axis, while dashed lines show the molecular gas mass (purple) and dust mass (green); using the scale on the right $y$-axis. Each galaxy is at a different stage of quiescence, ranging from more recently quenched in the left panel to earlier quenched in the right panel. The thin grey vertical dotted line marks the selection time, with earlier times to the right and later times to the left. The red vertical dotted line marks the corresponding quenching timescale (defined in sec.~3.2 of \citetalias{chandro-gomez_inprep}). The arrows indicate the snapshots at which we generate galaxy images for these systems (in addition to the $z=2$ snapshot, corresponding to the first value on the $x$-axis) in Figs.~\ref{fig:images0}--\ref{fig:images2}. The text in the top left corner indicates whether the galaxy is rejuvenated or non-rejuvenated at $z=2$.}}
\label{fig:images-h} 
\end{figure*}

\textcolor{black}{Fig.~\ref{fig:images-h} shows three panels, each corresponding to a different MQG selected at $z=3$ from the L200m6 simulation. Each panel presents the evolutionary history of several galaxy properties as a function of lookback time (with the corresponding redshift shown on the top $x$-axis). We show the SFR in red and the instantaneous BHAR \textcolor{black}{($\dot{M}_{\rm accr}$ for the thermal AGN feedback model)} in blue, with labels on the left y-axis. The molecular hydrogen mass (total $\rm H_2$ mass in gas particles within the fiducial aperture defined in \S~\ref{sec:prop}) and dust mass (total dust mass in gas particles) are shown as dashed purple and green lines, respectively. The corresponding TrackID for each system, as identified by {\sc HBT-HERONS}, is indicated in the top-right corner for reference.}

\begin{figure*}
\centering
\includegraphics[width=0.98\textwidth]{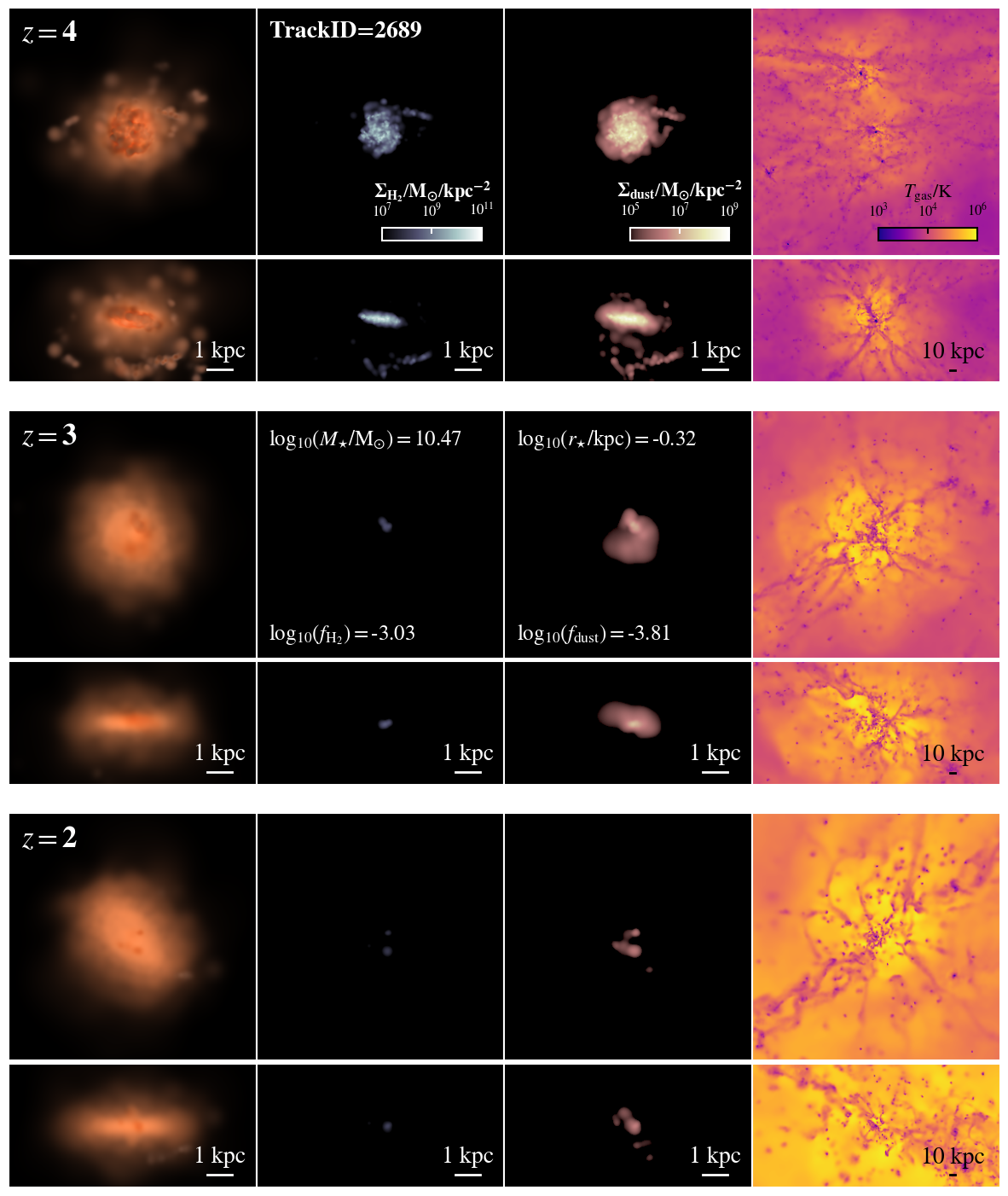}
\caption{\textcolor{black}{Galaxy images for the TrackID=2689 MQG presented in the left panel of Fig.~\ref{fig:images-h}. Each row shows the galaxy at a different stage of its evolution: $z=4$ (\textit{top}), the selection redshift $z=3$ (\textit{middle}), and $z=2$ (\textit{bottom}). For each epoch, we show both face-on and edge-on projections of the system, centred on the most bound particle of the subhalo (subhalo centre) and aligned with the stellar angular momentum vector computed within a spherical aperture of $10,\mathrm{pkpc}$. Each column corresponds to a different galaxy property: stellar light in \textit{JWST} colours including dust attenuation (\textit{first}); mean $\mathrm{H_{2}}$ surface density (\textit{second}); mean dust surface density (\textit{third}); and mass-weighted mean gas temperature (\textit{fourth}). The panels in the first three columns span $10,\mathrm{pkpc}$ across (and in thickness), while the panels in the fourth column span $0.5,\mathrm{pMpc}$ across (and in thickness). Galaxy properties at the selection time are indicated in the middle row panels.}}
\label{fig:images0} 
\end{figure*}

\begin{figure*}
\centering
\includegraphics[width=0.98\textwidth]{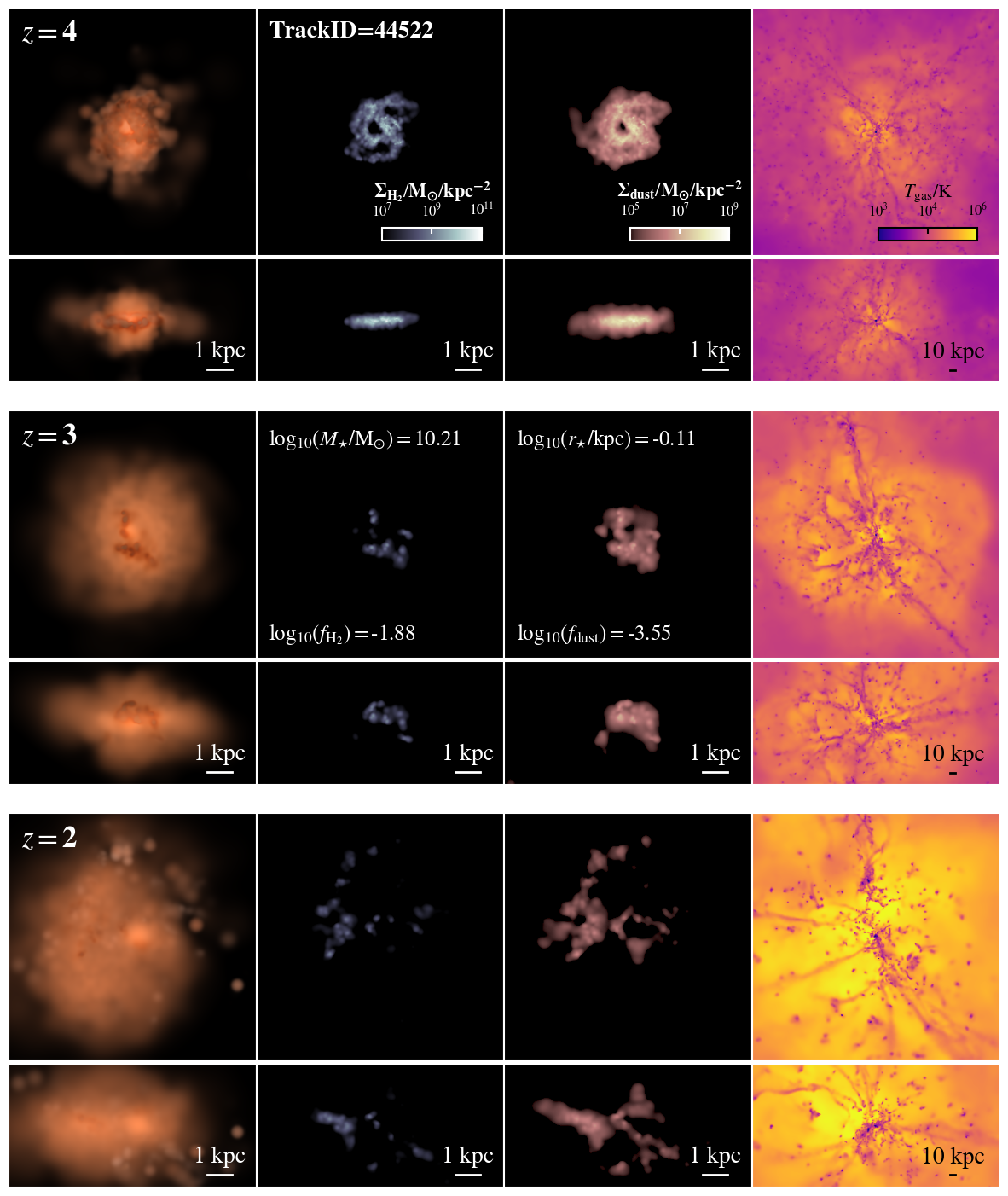}
\caption{\textcolor{black}{Same as Fig.~\ref{fig:images0}, but for the $z=3$ TrackID=44522 MQG shown in the middle panel of Fig.~\ref{fig:images-h}.}}
\label{fig:images1} 
\end{figure*}

\begin{figure*}
\centering
\includegraphics[width=0.98\textwidth]{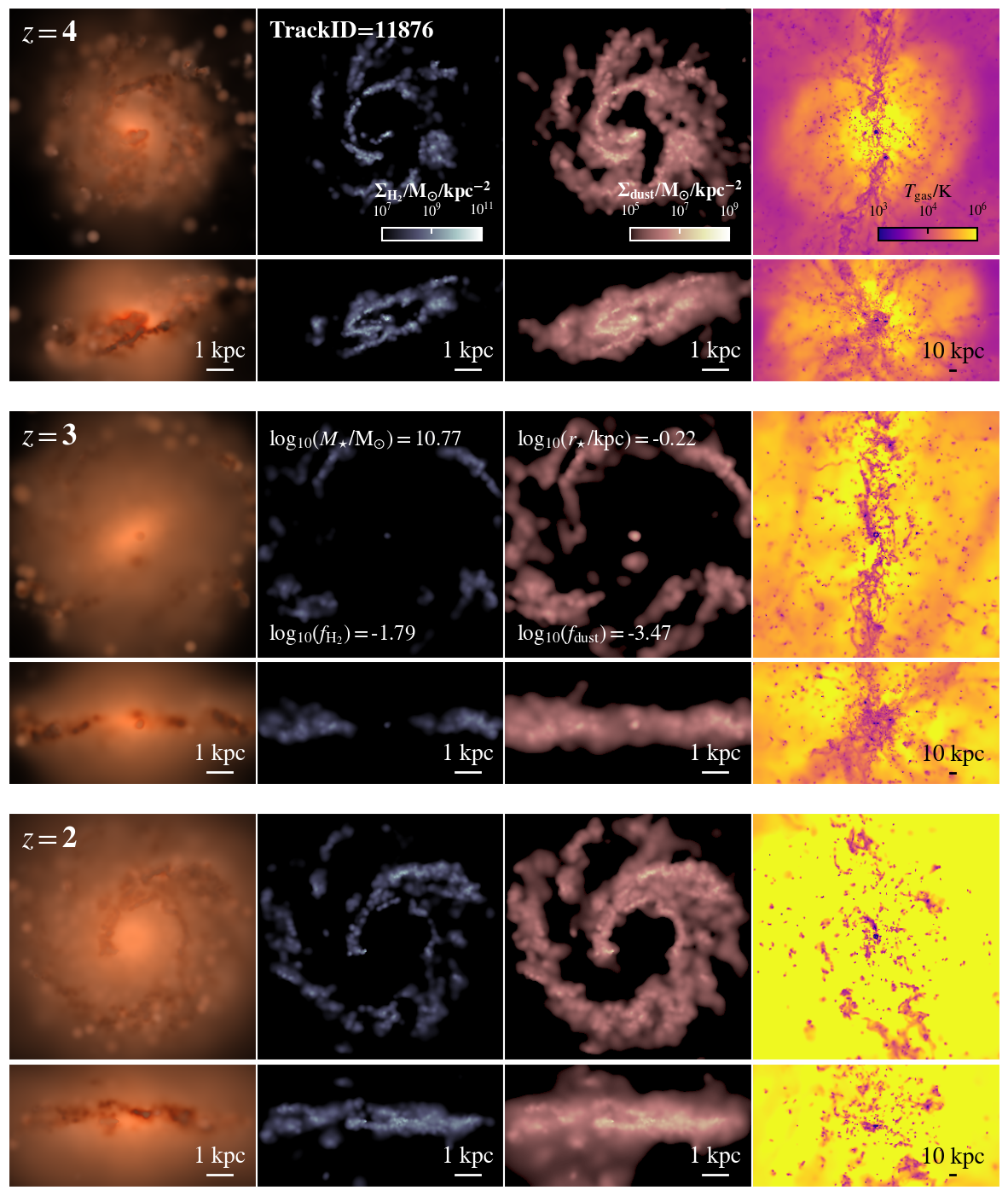}
\caption{\textcolor{black}{Same as Fig.~\ref{fig:images0}, but for the $z=3$ TrackID=11876 MQG shown in the right panel of Fig.~\ref{fig:images-h}.}}
\label{fig:images2} 
\end{figure*}

\textcolor{black}{We also include galaxy images for the same systems in Figs.~\ref{fig:images0}--\ref{fig:images2}. The stellar-light panels are produced using the software described in \citet{husko_inprep}, an approximate line-of-sight radiative transfer tool that neglects scattering into the line of sight \textcolor{black}{\citep[see section~5 of][for further details]{schaye25}}. In this case, three \textit{JWST} filters are combined to produce RGB images. The remaining images are generated using the {\sc Swiftgalaxy}\footnote{\url{https://github.com/SWIFTSIM/swiftgalaxy}} and {\sc Swiftsimio}\footnote{\url{https://github.com/swiftsim/swiftsimio}} Python packages. The three rows in these images correspond to $z=4$ (before selection), $z=3$ (selection), and $z=2$ (after selection), from top to bottom. These redshifts are marked with black vertical lines in the panels of Fig.~\ref{fig:images-h}. The panels show both face-on and edge-on projections based on the stellar angular momentum within a 10 pkpc spherical aperture. Each column displays a different property: stellar light in \textit{JWST} colours, including dust attenuation (first column), mean $\mathrm{H}_2$ surface density (second), mean dust surface density (third), and mass-weighted mean gas temperature (fourth). Colour bars indicate the range sampled by the last three quantities.}

\textcolor{black}{In Fig.~\ref{fig:images-h}, the selection time for each galaxy is marked by a vertical grey dotted line, while the quenching time, defined in sec.~3.2 of \citetalias{chandro-gomez_inprep}, is indicated by the vertical red dashed lines. Each system represents a different evolutionary stage, with galaxies in the right panels having quenched earlier. In all cases, a strong peak in the SFR occurs shortly before quenching (given the definition of quiescence we use). We also find that the BHAR generally peaks immediately before the quenching event, indicating that rapid BH growth leads to substantial energy release through AGN feedback (given the prescriptions described in \S~\ref{ssec:colibre-agn}). The energy released through AGN feedback is directly linked to BH growth; thus, BHAR closely traces the feedback energy injection. This provides further evidence that AGN feedback is responsible for quenching these systems (\S~\ref{ssec:res2-agn-feedback}).}

\textcolor{black}{In all three galaxies, the molecular gas content closely follows the SFR evolution, as expected since star formation occurs within dense, cold gas regions that would be dominated by molecular gas (as shown in \citealt{lagos25b}). SFR peaks correlate with large molecular gas and dust reservoirs, since dust facilitates the formation of $\rm H_2$, and both quantities are therefore naturally linked. Drops in SFR and BHAR are accompanied by declines in both molecular gas and dust masses, consistent with AGN feedback destroying or expelling the cold gas and dust reservoirs. \textcolor{black}{Because these galaxies are extremely compact, changes in the gas supply remain tightly coupled to the BHAR.} Additional analysis of dust and molecular hydrogen properties is presented in \citetalias{chandro-gomez_inprep} (sec.~4.4), where MQGs are shown to host significantly lower $\rm H_2$ and dust reservoirs than MSFGs. The galaxy images illustrate this behaviour directly, showing the disappearance of molecular gas (second column) and dust (third column) after quenching, together with AGN-driven hot bubbles clearly visible in the gas temperature distribution (fourth column), associated with strong feedback episodes.}

\textcolor{black}{The three galaxies nevertheless exhibit different post-quenching evolutionary paths. The TrackID=2689 system (left panel) shows a recently quenched galaxy in which AGN feedback has depleted or heated the molecular gas and reduced the dust reservoir. The dust-to-gas ratio ($\mathrm{DGR}=M_{\rm dust}/M_{\rm H_2}$) is relatively high, finding that systems with high DGRs (studied in \citetalias{chandro-gomez_inprep}) correspond to galaxies where molecular hydrogen is depleted more efficiently than dust, without evidence in those cases for dust regrowth in the absence of molecular gas recovery. Examples of galaxies that quenched earlier and where molecular hydrogen evolves more abruptly than dust are visible in the middle and right panels. The TrackID=44522 galaxy (middle panel) later regrows molecular hydrogen, while the TrackID=11876 galaxy (right panel) gradually recovers both significant dust and molecular gas. The images also show dust remaining in the outskirts of the 11876 system, which may appear surprising but is increasingly supported by recent observations \citep[e.g. GS-9209 galaxy in][]{deugenio26}. \textcolor{black}{Same with the extended emission of molecular gas, as in \citet{valentino26}}. In both cases (44522 and 11876), the systems eventually recover the gas content, but not enough to return to the star-forming population at $z=3$.}

\textcolor{black}{However, by $z=2$, the 44522 and 11876 galaxies experience rejuvenation episodes, while 2689 remains quenched. The $z=3$ images (middle row) show perturbations in the filamentary structure around the galaxies (fourth column), with inflowing cooler gas being disrupted by AGN-driven bubbles. In the case of 2689, this appears sufficient to suppress further gas supply by $z=2$. In contrast, for the other two systems (44522 and 11876), despite similar (and in the latter case more extreme) perturbations of the filaments, they are still able to retain or reaccrete some material from the surrounding medium, leading to renewed star formation. By $z=1$, all three systems are quenched again. \textcolor{black}{Observationally, \citet{gangula26} reported evidence for rejuvenation occurring $\sim500\,\mathrm{Myr}$ ago in a gravitationally lensed MQG, likely triggered by external gas accretion.}}

\bsp	
\label{lastpage}
\end{document}